# Tractography derived quantitative estimates of tissue microstructure depend on streamline length: A characterization and method of adjustment.

Richard G. Carson[a,b]  and Alexander Leemans[c]

[a]Trinity College Institute of Neuroscience and School of Psychology, Trinity College Dublin, Dublin 2, Ireland

[b]School of Psychology, Queen's University Belfast, Belfast, Northern Ireland, BT7 1NN, UK.

[c]Image Sciences Institute, University Medical Center Utrecht, Utrecht 85500, The Netherlands.

Corresponding Author:

Richard G. Carson

Trinity College Institute of Neuroscience and School of Psychology

Trinity College Dublin

Dublin 2,

Ireland,

E-mail: richard.carson@tcd.ie

**Abstract**

Tractography algorithms are used extensively to delineate white matter structures, by operating on the voxel-wise information generated through the application of diffusion tensor imaging (DTI) or other models to diffusion weighted (DW) magnetic resonance imaging (MRI) data. We demonstrate that these methods commonly yield systematic streamline length dependent distortions of tractography derived tissue microstructure parameters, such as fractional anisotropy (FA). This dependency may be described as piecewise linear. For streamlines shorter than an inflection point (determined for a group of tracts delineated for each individual brain), estimates of tissue microstructure exhibit a positive linear relation with streamline length. For streamlines longer than the point of inflection, the association is weaker, with the slope of the relationship between streamline length and tissue microstructure differing only marginally from zero. As the dependency is most pronounced for a range of streamline lengths encountered typically in DW imaging of the human brain (less than ~100 mm), our results suggest that some previous estimates of tissue microstructure should be treated with considerable caution. A method is described, whereby an Akaike information weighted average of linear, Blackman and piecewise linear model predictions, may be used to compensate effectively for the dependence of FA (and other estimates of tissue microstructure) on streamline length, across the entire range of streamline lengths present in each specimen.

**Keywords**

**1. Introduction**

Tractography algorithms are used extensively to delineate white matter structures, by operating on the voxel-wise information generated through the application of diffusion tensor imaging (DTI) or other models to diffusion weighted (DW) magnetic resonance imaging (MRI) data. With the aim of delineating only those tracks (consisting of "streamlines") that are anatomically plausible, tractography algorithms incorporate certain criteria. For example, a curvature threshold (e.g., between 30° and 70°) may be applied to exclude deviations that would otherwise result in propagation to an adjacent streamline. Magnitude thresholding on fractional anisotropy (FA) values (or on the fibre orientation distribution (FOD) in the case of the constrained spherical deconvolution (CSD) model) results in the termination of tracking when the value for a voxel falls below the defined threshold (e.g., FA < 0.20).

In the present paper, we demonstrate that these methods yield systematic streamline length dependent distortions of tractography derived tissue microstructure parameters, such as FA. These effects arise as FA or FOD values for successive voxels defining a streamline do not exhibit an abrupt transition from values well in excess of the magnitude threshold, to a value falling below the threshold. Rather, the distribution of values obtained for such measures tends to be relatively smooth, with the largest values being obtained close to the middle of the streamline, and the lowest values (by definition) at the two ends of the streamline (Zhang et al., 2018). It follows that the relative contribution of "low-values" to the total estimate will be larger for shorter streamlines (and fibre tracks) than for longer streamlines. We show that these factors give rise to a clear and consistent dependency of the estimated FA values on streamline length.

It is not our purpose to engage in detailed biophysical modelling of this previously unreported phenomenon. Rather, our objectives are to 1) demonstrate that the effect is clear and in practical terms far from trivial, and 2) outline the relatively straightforward steps that may be taken to compensate for its impact on tractography derived estimates of tissue microstructure.

## 2. Methods

With a view to illustrating the key features of this phenomenon, we first present a re-analysis of data acquired in work presented by Ruddy et al. (2017). Analyses of additional data are described in the Supplementary Material. The relevant features of the data acquisition and pre-processing undertaken by Ruddy et al. (2017) are as follows.

The participants were forty-three neurologically healthy right-handed volunteers (aged $22.5 \pm 2.9$ SD, 28 female). All gave informed consent to procedures that (with the exception of preregistration) were in accordance with the Declaration of Helsinki. These had been approved by the appropriate Queen's University Belfast and Trinity College Dublin Ethics Committees.

A 3T Philips Achieva magnetic resonance scanner, with an eight-channel head coil, was used to acquire diffusion weighted images. The sequence comprised single shot echo planar imaging (EPI) with a slice thickness of 2.29 mm, repetition time = 9994 ms, echo time = 73 ms, number of diffusion directions = 61, b value = 1500 s/mm$^2$, number of slices = 60 (transverse), in-plane resolution $2.3 \times 2.3$ mm$^2$, with a field of view of 258 mm (RL) $\times$ 258 mm (AP) $\times$138 mm (FH).

ExploreDTI (Leemans et al. 2009) was used for data processing. Images were corrected for head movement and eddy currents using the procedure described in Leemans and Jones (2009). Tensor estimation was performed using the iteratively reweighted linear least squares approach (Veraart et al. 2013). Fibre trajectories were computed with CSD based tractography (Tournier et al. 2007). When compared to conventionally applied DTI based fiber tractography (FT), this method increases the sensitivity with which functionally significant variations in white matter characteristics may be detected (Reijmer et al., 2012). Recursive calibration of the response function was used to optimise the estimation of the fibre orientation distribution (FOD) functions (Tax et al. 2014). A uniform grid of tractography seed points at a resolution of $2 \times 2 \times 2$ mm$^3$ was employed, with an angle threshold of 30 degrees, an FOD threshold of 0.1, and a maximum harmonic order of eight.

The cortical motor network atlas developed in Ruddy et al. (2017) was used, including the following regions: posterior and anterior primary motor cortex (M1a and M1p), dorsal and ventral premotor cortex (PMd and PMv), supplementary motor area proper (SMA proper) and pre-supplementary motor area (pre-SMA), primary sensory cortex (S1), and the cingulate motor area (CMA) in both hemispheres. Reconstructed fibre trajectories for all pairwise combinations of brain regions were quantified separately for each individual participant. As a result, a total of 120 tracts were obtained: 64 transcallosal tracts, and 28 tracts within each hemisphere. We herein focus on the relationship between FA and streamline length. Analyses of AFD, radial diffusivity (RD) and mean diffusivity (MD) are described in the Supplementary Materials.

## 3. Results

As it is known that distributions of FA values (and indeed all eigenvalue-based measures (Babamoradi et al., 2013)) deviate from normality (Cascio et al. 2013; Clement-Spychala et al., 2010), robust statistical methods were used throughout. All reported confidence intervals (c.i.) are bias corrected adjusted (bca), based on 1000 bootstrap samples. In respect of correlations, these were converted to z scores before bootstrap resampling (Gorsuch & Lehmann, 2010), the bca statistics were weighted by sample size (Karyawati et al., 2020), and the inverse transform then applied.

As a first step we calculated the Kendall rank correlation coefficient, to characterise the ordinal association of FA and streamline length. The magnitude of this coefficient expresses the similarity of the orderings of the two samples. This was done in two ways. In the first, the coefficient was calculated separately for each of the 43 participants. As streamlines were not resolved for all of the 120 tracts, the mean number of tract observations included in each calculation was 82 (95% c.i. 80-83). The mean value of tau – the Kendall correlation coefficient, was then obtained across participants. The mean magnitude of the correlation was 0.35 (95% c.i. 0.30 – 0.39), corresponding to an effect of "moderate" size. In other words, within individual brains, tracts with longer streamlines are characterised by larger FA values. In the second method of analysis, the coefficient was calculated separately for each of the tracts (e.g., left M1a to right M1a), using the sample of 43 participants. The mean value of tau across all tracts was then derived. In this case, each bootstrapping sample comprised a random selection of tracts (rather than participants). The mean magnitude of the correlation was 0.28 (95% c.i. 0.24 – 0.31). Thus, when any given tract is considered across different brains, individuals with longer streamlines tend to exhibit larger FA values.

It should not however be assumed that the magnitude of the dependency of FA (or AFD etc) on streamline length remains constant across all streamline lengths. A simple model derived from basic assumptions illustrates this point (Figure 1). We emphasise that this is a statistical, rather than a biophysical, model. It encompasses the reality of magnitude thresholding, whereby tracking is terminated when the value obtained for a voxel falls below a pre-defined threshold. It also includes the empirical observation that estimates of tissue microstructure diminish progressively towards the ends of a streamline (Zhang et al., 2018), It can further be assumed that these estimates are constrained to a finite range (e.g., 0 to 1).

Given this set of assumptions, it is predicted that average (or median) estimates of tissue microstructure (i.e., for an entire streamline) will increase in a linear fashion with extensions of streamline length (Figure 1, Panels A to J), up to the length at which the upper limit of the range of possible values (e.g., 1) is reached. For streamlines that exceed this length (Figure 1, Panels K to T), the average estimate of tissue microstructure (for an entire streamline) will increase as a power function with further increments in streamline length (Figure 1, Panel U). But a key caveat applies. Real biological specimens do not yield values with a magnitude equal to the upper limit of the potential range. Rather, for each specimen (i.e., for each individual brain) there will exist an asymptotic value for the derived estimate of tissue microstructure (Figure 1, Panel V). At streamline lengths beyond that at which the asymptote is obtained, no further increase in the average estimate of tissue microstructure (FA, AFD etc), i.e., no dependency on streamline length, is to be expected.

In line with this characterisation, non-linear curve fitting using the Blackman (also known as a linear-plateau) function (Archontoulis & Miguez, 2015) (Figure 2) to characterise the relationship between FA and streamline length indicated that in 41 of the 43 participants, the fit to the Blackman function was superior to that achieved using a linear model. It is however an intrinsic feature of the Blackman response that the slope is constrained to be zero at streamline lengths beyond that of the asymptote. To accommodate the possibility that the slope obtained empirically differs from zero, a piecewise linear model can therefore also be applied. The modelling steps described were undertaken in R (R Core Team, 2019), through quantile regression, using the rq() and nlrq() functions provided in the "quantreg" package (Koenker et al., 2018). The SSlinp() function in the "nlraa" package (Miguez et al., 2021) was used as a self-starter for the coefficients of the Blackman function. Piecewise linear models were fitted using the "segmented" package (Muggeo, 2021).

One of our goals was to derive a means of compensating for the impact of streamline length on estimates of tissue microstructure. To this end, we employed model-averaging. It has been demonstrated that model-averaged estimators improve precision and reduce bias, relative to estimators derived from a selected "best model" (Burnham and Anderson, 2002). When the aim is prediction and the calculation of residuals – which is central to the present undertaking, model averaging is always superior to a best-model strategy (Burnham & Anderson, 2004). Recognising that any model is merely an approximation to the truth, Akaike's information criterion (AIC) was used to quantify the three candidate models (linear, Blackman, and piecewise-linear) in terms of information loss. In practice, the modified version (AICc), that is suitable for small sample sizes, was used. An Akaike weight was then calculated for each model (i.e., separately for each data sample/participant). The Akaike weight takes a value between 0 and 1. Importantly, the weights of all models in a candidate set sum to 1. The Akaike weight is analogous to the probability that any given model is the best approximating model of the data (Symonds & Moussalli, 2011). A weighted average of model predictions is obtained by multiplying the predicted values generated by each model by the corresponding Akaike weight, and then summing across the (in this case three) sets of weighted predicted values (Burnham & Anderson, 2002; Symonds & Moussalli, 2011). A model with a low Akaike weight thus has little influence on prediction (Figure 3A).

The model-averaging approach can also be applied to the individual parameter estimates generated by candidate models. The Blackman and the piecewise-linear models yield estimates of the inflection point (or "breakpoint") that marks the termination of the initial segment. In addition, both models provide estimates of the slope of the initial segment, and of the slope of the second segment (which for the Blackman model is equal to zero). Model averaged estimates for these parameters were obtained in the manner described above (the Akaike weights were rescaled such that with the exclusion of the linear model the sum remained equal to 1).

The mean value of the inflection point (i.e., the transition between the two segments for the Blackman and piecewise models) calculated across participants was 102.1 mm (95% c.i. 98.6 – 104.8 mm). The confidence interval (c.i. = 6.2 mm) obtained for the inflection point was considerably smaller than that of the range of streamline lengths (c.i. = 26.0 mm) observed across participants. The mean value of the FA obtained at the inflection point was 0.45 (95% c.i. 0.44 – 0.46). The slope of the first segment was 0.0041 (95% c.i. 0.0034 – 0.0050). In other words, each 10 mm increase in streamline length gave rise to an increase in FA of approximately 0.04. This indicates that over the range of streamline values spanned by the initial segment (i.e., up to approximately 102 mm), the FA estimate depends on the length of the streamline for which it was derived. Consistent with this supposition, the Kendall tau values of correlations between streamline length and FA, calculated for streamlines shorter than that of the inflection point (derived separately for each individual) corresponded to large effect sizes (mean = 0.51, 95% c.i. 0.44 – 0.55). It is very clearly the case therefore that below the inflection point, streamline length has a substantial determining influence on FA estimates (Figure 4).

In contrast, the mean slope of the second segment (Figure 4) did not differ reliably from zero (mean = -0.0006, 95% c.i. -0.0014 – 0.00008). In this case, a 10 mm increase in streamline length was associated with a decrease in FA of 0.006. The correlation between streamline length and FA in this region was similarly weak (mean = -0.09, 95% c.i. -0.15 – -0.013).

The model-averaged (predicted) values represent the optimal fit to the relationship between streamline length and the estimate of tissue microstructure, at the values of streamline length present in the data available for each individual participant. The residuals (Figure 3B) correspond to the variations in the estimates of tissue microstructure that are not determined by streamline length. A positive value of the residual indicates that, for a given tract, the estimate of tissue microstructure is larger than that which would be predicted by streamline length alone – to an extent corresponding to its magnitude. A negative residual value for a tract indicates an estimate of tissue microstructure smaller than would be predicted for its streamline length. The mean magnitude of the Kendall correlation between the residuals and streamline length, when calculated across participants, was -0.009 95% (c.i. -0.022 – 0.005). Calculated across tracts, the correlation between the residuals and streamline length did not differ reliably from zero (mean = -0.003, 95% c.i. -0.054 – 0.045). The model averaging approach was therefore successful in compensating for the influence of streamline length on FA.

The use of the residuals in inferential analyses may prove to be sufficient in circumstances in which the research question can be addressed by dealing with variations in the *relative magnitude* of the estimates of tissue microstructure obtained for a defined set of tracts within each brain. There are however many instances in which it is desirable to compare the estimates of tissue microstructure obtained for a specific tract, across groups of individuals, or to examine changes that occur in individuals over time. To address this requirement, it is necessary that the residuals are expressed relative to an appropriate reference value. The model-averaging approach affords a means of addressing this requirement. Specifically, the model averaged estimate of FA (or AFD etc) at the inflection point provides an appropriate, individual specific, reference value. The magnitudes of the estimates obtained in the present analysis (median of 0.45 for FA across all individuals) have face validity. In so much as the fitted values of the model do not tend to vary appreciably over the range of streamline lengths extending beyond the inflection point (as indicated by the negligible slopes of the second segment), the model averaged estimate of tissue microstructure at the inflection point can reasonably be taken as an asymptotic value. Thus, it also has construct validity. It is therefore proposed that the adjusted (for streamline length) measure be obtained (i.e., separately for each brain) as the sum of the model averaged estimate of FA (or AFD etc) at the inflection point, and the residual derived for each tract.

The adjusted values derived in the manner described, differ systematically from the original FA values that are analysed routinely in tractography studies. Necessarily the magnitude of the difference varies as a function of streamline length (Figure 5). Estimates generated for tracts with the shortest streamlines (typically < 70 mm) are characterised by adjusted FA values that exceed the original values by more than 0.1. In respect of tracts with streamline lengths greater than the mean inflection point of 102 mm (estimated across participants), the magnitude of the difference tends to be smaller, and frequently cannot be distinguished reliably from zero (Figure 5).

The broader implications of the distortions in FA values that arise from streamline length dependence are readily demonstrated. The first panel of Figure 6 (A) illustrates the rank ordering of unadjusted FA values obtained for streamlines (n = 26 tracts) that originate and terminate within the right cerebral hemisphere. The second panel (B) illustrates the rank ordering of FA values that have been adjusted for streamline length in the manner described above. It is apparent that the rank ordering of the adjusted values is dramatically different from that of the unadjusted values. This is reflected in a correlation of the rankings for the unadjusted and unadjusted FA values that is markedly lower than unity (Kendall's tau = 0.34).

It is not difficult to envisage specific research questions for which the outcomes may depend on whether the data have first been adjusted to compensate for the influence of streamline length. For example, an investigator may wish to determine whether tracts that originate and terminate within one hemisphere differ from "inter-hemispheric" tracts, with respect to a tissue microstructure parameter such as FA. For the data set described, the central tendency (derived using trimmed means) of the unadjusted FA values (Figure 7, Panel A) obtained for "intra-hemispheric" tracts within the right hemisphere is 0.404, whereas, for "inter-hemispheric" tracts connecting the left and right hemispheres, the corresponding value is 0.427 (difference = -0.022, 95% c.i. -0.031 – -0.015). The outcomes of a robust inferential test (Yuen's t) appropriate for paired observations (i.e., intra-hemispheric versus inter-hemispheric) indicate that this is a large (in terms of standardised effect size) and reliable effect (t(26) = -5.17, p = 2.13E-05, $\delta t$ = -0.59, 95% c.i. -0.81 – -0.39). In marked contrast, the difference between the adjusted FA values (Figure 7, Panel B) obtained for the intra-hemispheric (0.448) and inter-hemispheric (0.453) tracts is -0.004 (95% c.i. -0.009 – 0.0008). The associated inferential test (t(26) = -1.29, p = 0.175, $\delta t$ = -0.08, 95% c.i. -0.22 – 0.01) supports the conclusion that, when adjusted for streamline length, the FA values obtained for streamlines defined for intra-hemispheric tracts do not differ reliably from those defined for inter-hemispheric tracts.

The model-averaging approach described is based on determining the optimal fit to the relationship between streamline length and the estimate of tissue microstructure, at the values of streamline length present in the data available for each person. It is also possible to model the relationship between streamline length and the estimate of tissue microstructure, using the data available for each tract. In many instances however, this approach will be insufficient to provide an adequate characterisation of the relationship. This is because the extent of the variation across individuals in the length of the streamlines identified for a specific tract, will typically be much smaller than the range of the streamline lengths that characterise the entire set of tracts defined for each person. For the data considered herein, the mean (i.e., across 43 individuals) of the range of streamline lengths detected for all tracts (defined using the cortical motor network atlas) was 106.4 mm. In contrast, the mean for all tracts, of the range of streamline lengths obtained for each tract, was 63.9 mm. Necessarily also, the range of streamline lengths obtained (across persons) for a given tract will span only a subset of the range of streamline lengths defined for any given person.

The impact of these particulars on the data modelling is illustrated by Figure 8. In Panel A, the left cingulate motor area to left dorsal premotor cortex tract (PMd) is represented. For 39 of the 43 individuals, the length of the streamlines defined for this tract was shorter than the breakpoint defined for the FA values (102.1 mm). For this range of streamline lengths, the relationship with FA is predominantly linear. In Panel B, the left CMA to right anterior primary motor cortex (M1a) tract is represented. For 18 of the 42 individuals for whom streamlines were detected, these were shorter than the FA breakpoint. Whereas, for 24 individuals, the streamlines were longer than the breakpoint. In this case, a piecewise linear relationship is apparent. In Panel C, the left PMd to right PMd tract is shown. For all 43 individuals, the lengths of the streamlines detected for this tract exceeded the breakpoint defined for FA. With respect to this tract, there is no apparent relationship between streamline length and FA. These examples serve to highlight that, an adequate characterisation of the relationship that exists between streamline length and this estimate of tissue microstructure, demands a range of streamline lengths beyond that which is available for a single tract.

## 4. Discussion

We have demonstrated that there is a clear and robust dependence of FA on the length of the streamlines for which this measure of tissue microstructure is derived. This dependency is described well as piecewise linear (Figure 4). In our illustration, we showed that for a range of streamline lengths below a point of inflection, each 10 mm increase in streamline length was associated with an increase in FA of approximately 0.04 In other words, if two tracts both with streamlines shorter than 100 mm or so, vary in their respective streamline lengths by 25 mm, there will tend to be a difference of approximately 0.1 in the magnitude of their FA estimates. For streamlines longer than the point of inflection, the association is very much weaker, with the slope of the relationship between streamline length and AFD differing only marginally from zero.

It was illustrated (Figure 1) that an influence of streamline length on estimates of tissue microstructure can arise if FA values (or other estimates of tissue microstructure) diminish progressively towards the ends of a streamline (Zhang et al., 2018), and a fixed threshold is used (i.e., by a tractography algorithm) to terminate tracking. The specific relationship between streamline length and FA observed empirically (Figure 4) arises from the fact that in real brains, the estimate of tissue microstructure derived for any given specimen will be less than the theoretical upper limit defined for that measure (Figure 1, Panel V). As the observed relationship appears therefore to have a principled basis, it seems probable that it will be ubiquitous (see Supplementary Materials). Given that the dependency is most pronounced for a range of streamline lengths encountered typically in (human brain) DW imaging (i.e., shorter than 100 mm or so), many previous estimates of tissue microstructure should be viewed with extreme caution. Indeed, it is only tracks composed of streamlines with lengths that exceed the inflection point, for which the estimates are likely to be veridical.

Due to the piecewise linear nature of the relationship between streamline length and the estimate of tissue microstructure, it will not generally be feasible to adjust for its influence by simply including streamline length as a linear covariate in downstream analyses (e.g., Vos et al., 2011). There may be exceptions, if for example it can be established that the range of streamline lengths for all tracts is below the inflection point defined for the specimen in question. There is however a more secure alternative. We have outlined that an Akaike information weighted average of linear, Blackman and piecewise linear model predictions, may be used to compensate effectively for the dependence of FA on streamline length – across the entire range of streamline lengths present in each specimen. If the research question requires only consideration of the relative magnitude (within each brain) of the estimates of tissue microstructure obtained for a specific set of tracts, it may be sufficient to employ the residuals of the weighted average model predictions. In many cases however, it will be desirable to compare the estimates of tissue microstructure obtained for a specific tract across different samples. We have proposed that the estimated value of tissue microstructure obtained at the inflection point of the weighted average model, provides a basis upon which to adjust the residuals derived for each brain, in a manner sufficient to permit valid comparisons across samples. These considerations emphasise the importance of including as many tracts as possible in the modelling process, in order to best characterise the nature of the streamline length dependency. It is particularly desirable to include tracts with lengths sufficient to define the inflection point for each specimen. In practice therefore, it may be necessary to incorporate in the model, tracts that are not of direct interest in the analyses that follow.

## 5. Data and code availability statements

Example R code, which can be used to implement the methods of analysis described herein, is available via: https://zenodo.org. The Digital Object Identifier (DOI) is:

In respect of the brain imaging files upon which these analyses are based, three participants provided consent for their exemplar, pseudo-anonymised, data to be placed in the public domain. The example R code can be used to analyse these data, which are made available at the same location.

## 6. CRediT authorship contribution statement

**Richard G. Carson**: Conceptualization, Data analysis, Writing - Original Draft, Writing - Reviewing and Editing.

**Alexander Leemans**: Writing - Reviewing and Editing.

## 7. Funding

This research did not receive any specific grant from funding agencies in the public, commercial, or not-for-profit sectors.

## 8. Declaration of Competing Interests

The authors declare that they are not aware of any competing interests in respect of this work.

## 9. Acknowledgements

Supplementary Materials. The guidance provided by Fernando Miguez in relation to the deployment of model averaging is gratefully acknowledged.

## 10. Supplementary Material

Please see attached file.

**Figure captions**

1) For illustrative purposes, the theoretical estimate of tissue microstructure along a streamline is modelled as increasing with streamline length from a termination magnitude threshold of 0.2, at a rate equivalent to an increase of 1.6 for every change of 100 units of length. It is further assumed that the estimate of tissue microstructure is constrained to a range of 0 to 1. Given these parameters, the terminal sections (less than 50 units of length) – at both ends of the streamline, are characterised by all values being lower than 1. It is specified that in these regions the slope is constant, and independent of the overall length of the streamline. Within centre (non-terminal) sections, all values are equal to 1 (the theoretical asymptote). Average estimates of tissue microstructure (i.e., for an entire streamline) will therefore increase in a linear fashion with increases in streamline length (Panels A to J), up to the length at which the upper limit of 1 is reached (i.e., encompassing only the two terminal sections). For streamlines in excess of this length (i.e., also encompassing a centre section) (Panels K to T), the estimate of tissue microstructure will then increase as a power function with further increments in greater streamline length (Panel U). Crucially however, the tissue of real specimens will be characterised by estimates of tissue microstructure that are lower than the theoretical average. That is, the values obtained empirically will not continue to increase towards the theoretical limit (1) with increases in streamline length. In the present illustration, a maximum value of 0.7 is assumed. It follows that for this specimen the empirical average will remain 0.7 for all streamline lengths for which the theoretical model predicts a value of 0.7 or above. It would be anticipated therefore that the average estimate of tissue microstructure will increase in a linear fashion until the streamline length at which a value of 0.7 is obtained and remain constant (slope equal to zero) for streamlines in excess of this length (Panel V).

2) Theoretical fits to the Blackman (also known as a linear-plateau) function are illustrated. In this example, notional data from three specimens are shown to differ only with respect to the asymptotic value of the derived estimate of tissue microstructure. Specimen 1 (orange *) exhibits an asymptotic value of 0.72; Specimen 2 (blue ×) exhibits an asymptotic value of 0.80; and Specimen 3 (green +) exhibits an asymptotic value of 0.88. The streamline length corresponding to the start of the plateau region (the point of inflection) varies accordingly. In these examples, the slope of the initial segment is equivalent in each case. Empirical fits derived using the Blackman function can however also vary with respect to the slope of the initial segment. Furthermore, variations in the asymptotic value can covary with the streamline length at which the start of the plateau region occurs (Archontoulis & Miguez, 2015).

3) A. Empirical fits to each of the three candidate models (linear (red), Blackman (orange), and piecewise-linear (blue), and the model averaged fit (black), are shown for the 84 tracts delineated for a single individual. The Akaike weights for the respective models were as follows: linear – 0.007; Blackman – 0.564; piecewise-linear – 0.270. The point of inflection was determined to be 103.0 mm for the Blackman model (FA = 0.390), 102.0 mm for the piecewise-linear model (FA = 0.383), and 102.7 mm for the averaged model (FA = 0.388). The slope of the initial segment was estimated to be 0.003 for the Blackman model, the piecewise-linear model, and the averaged model.  The slope of the second segment was estimated to be 0.0002 for the averaged model. B. The FA values, adjusted for the influence of streamline length (through the application of a model averaging approach), are plotted for the same individual. In both panels, the colours are assigned to tracts in the order in which they are listed in the source data.

4) A summary representation of the results obtained through the application of a model averaging approach to the predicted FA values generated by the three candidate models (linear, Blackman, and piecewise-linear), for the 43 participants included in Ruddy et al. (2017). For each participant, the models were evaluated, and the predictions weighted and averaged, at a range of nominal streamline lengths (at 1mm intervals). This range spanned the median minimum streamline length and the median maximum streamline length observed across the 43 participants. The solid line corresponds to the means of the weighted, averaged, predicted values derived from 1000 bootstrapped samples. The dashed line was generated using the lower 95% confidence interval of the bootstrapped samples. The dotted line was generated using the upper 95% confidence interval. It is apparent that, for streamlines shorter than approximately 100 mm, there is a dependence of FA on streamline length.

5) Separately for each of the 43 participants included in Ruddy et al. (2017), and for each tract, the difference between the original FA value, and the FA value adjusted for the influence of streamline length (through the application of a model averaging approach) was calculated. The filled symbols correspond to the means of these difference values, when derived from 1000 bootstrapped samples drawn from the set of 43 participants (i.e., calculated separately for each tract). For a tract to be included, it was necessary that at least half of the participants must have contributed data (i.e., streamlines were resolved). The error bars correspond to the 95% confidence intervals of the bootstrapped samples. The tracts are plotted in order of mean streamline length (calculated across participants).

6) A. Rank ordering of unadjusted FA values obtained for streamlines (n = 26 tracts) that originate and terminate within the right cerebral hemisphere. B Rank ordering of FA values that have been adjusted for streamline length in the manner described in the text. The size of each symbol corresponds to the associated FA value (i.e., derived for that tract). The position of each symbol in relation to the y axis scale corresponds to the ranking of the FA value with respect to the set of 26 tracts. Those with a higher ranking (larger FA value) appear above those with a lower ranking (smaller FA value). While the assignation of fill colour to the individual tracts is arbitrary, as it remains consistent across panels A and B, it aids in the identification of differences in ranking (i.e., unadjusted (Panel A) versus adjusted (Panel B)). In Panel A, the colours are assigned to tracts in the order in which they are listed in the source data and plotted in this order in relation to the x axis scale. It is apparent that the rank ordering of the adjusted values (Panel B) is dramatically different from that of the unadjusted values (Panel A). M1a - anterior primary motor cortex; M1p - posterior and primary motor cortex; PMd - dorsal premotor cortex; PMv - and ventral premotor cortex; SMA proper - supplementary motor area proper; pre-SMA - pre-supplementary motor area; S1- primary sensory cortex; CMA - cingulate motor area.

7) The trimmed mean (trimming = 20%) of the FA values derived for all inter-hemispheric tracts for which streamlines were resolved for each participant, and the trimmed mean of the FA values derived for all right hemisphere intra-hemispheric tracts for which streamlines were resolved for each participant, were calculated. Unadjusted FA values are shown in Panel A. FA values that have been adjusted for streamline length in the manner described in the text are shown in Panel B. Each point represents the data for a single participant, with the colour coding determined by the order in which the 43 participants are listed in the source data. In both panels, the x coordinate for each point corresponds to the trimmed mean FA value for the right hemisphere tracts, and the y coordinate corresponds to the trimmed mean FA value for the inter-hemispheric tracts. Points lying close to the line of equality indicate similarity in the FA values obtained for inter-hemispheric and right hemisphere streamlines. Points lying below the line of equality indicate FA values for the right hemisphere streamlines that are lower than those for the inter-hemispheric streamlines. Comparison of the plots generated for the unadjusted data (Panel A) and the adjusted data (Panel B) makes apparent that differences between right hemisphere and inter-hemispheric FA values present for the unadjusted data, are not apparent for the adjusted data. The results of corresponding inferential tests are reported in the text.

8) Empirical fits to each of the three candidate models (linear (red), Blackman (orange), and piecewise-linear (blue), and the model averaged fit (black), are shown for three example tracts. Each point represents the data for a single participant, with the colour coding determined by the order in which the 43 participants are listed in the source data. A. Between left cingulate motor area (CMA) and left dorsal premotor cortex (PMd). B. Between left CMA and right anterior primary motor cortex (M1a). C. Between left PMd and right PMd.

## Figure 1

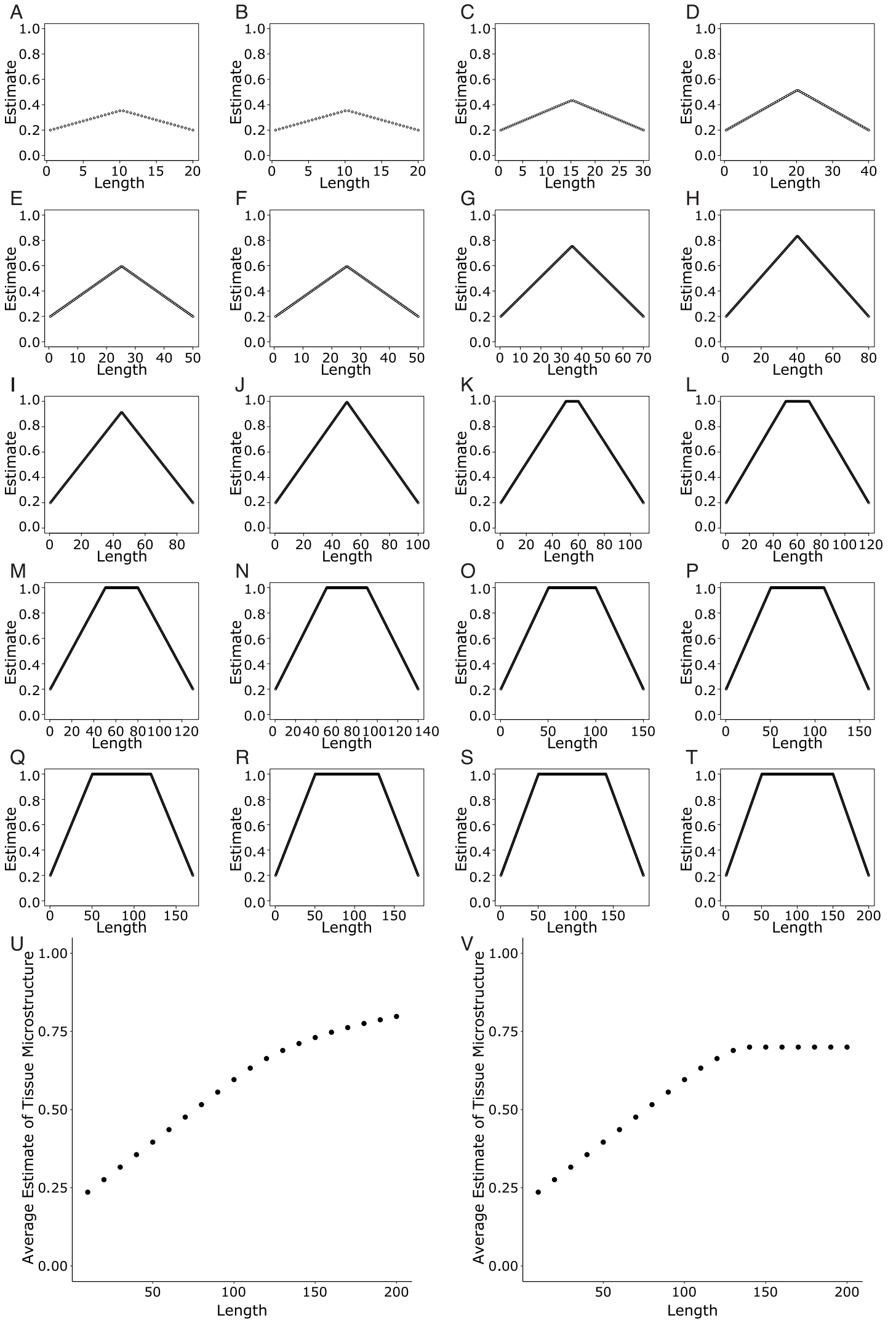

Figure 2

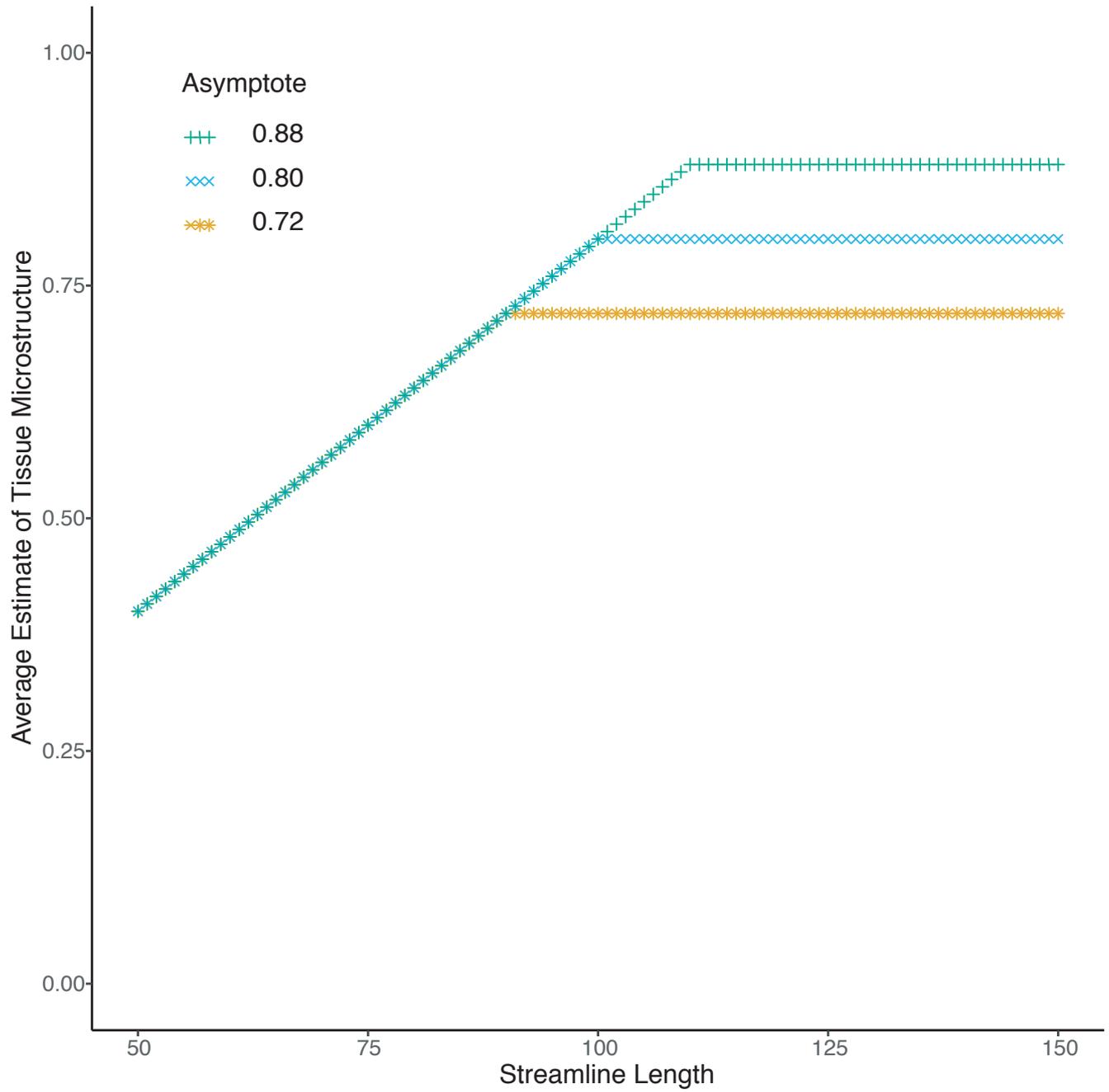

Figure 3

A

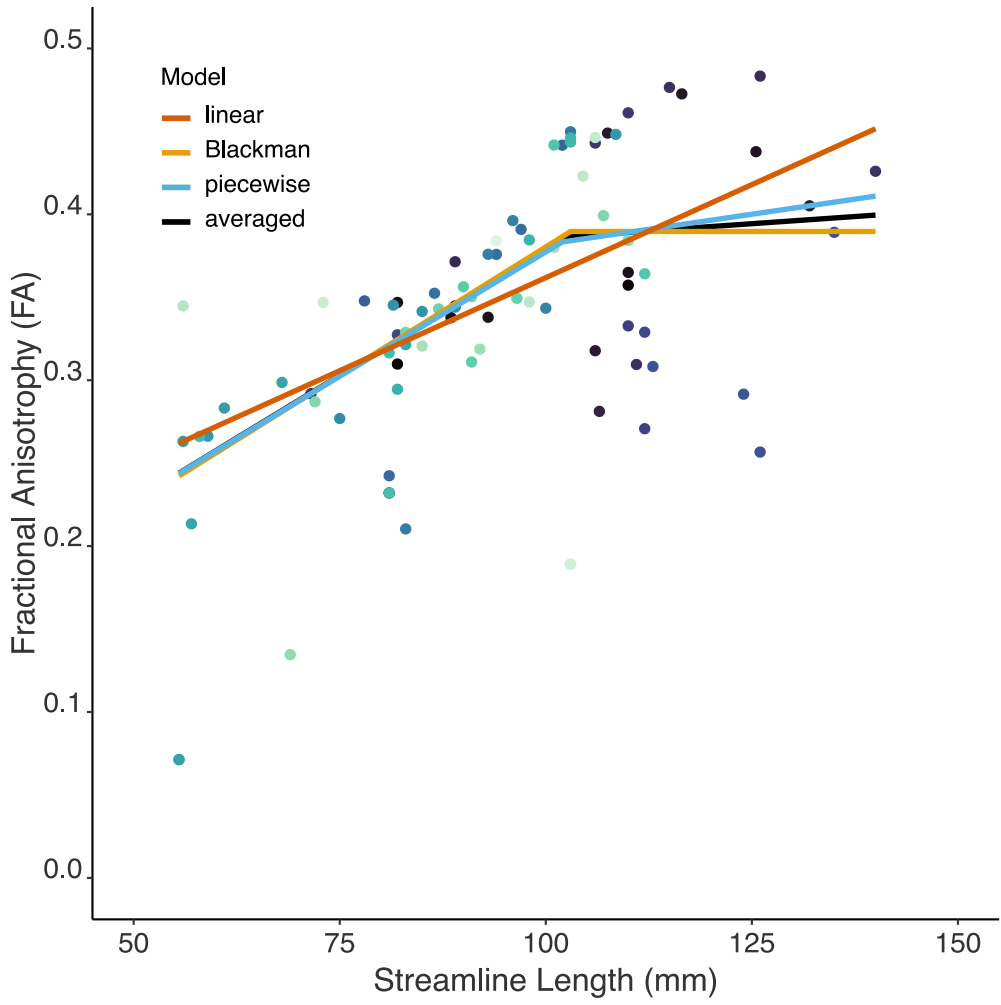

B

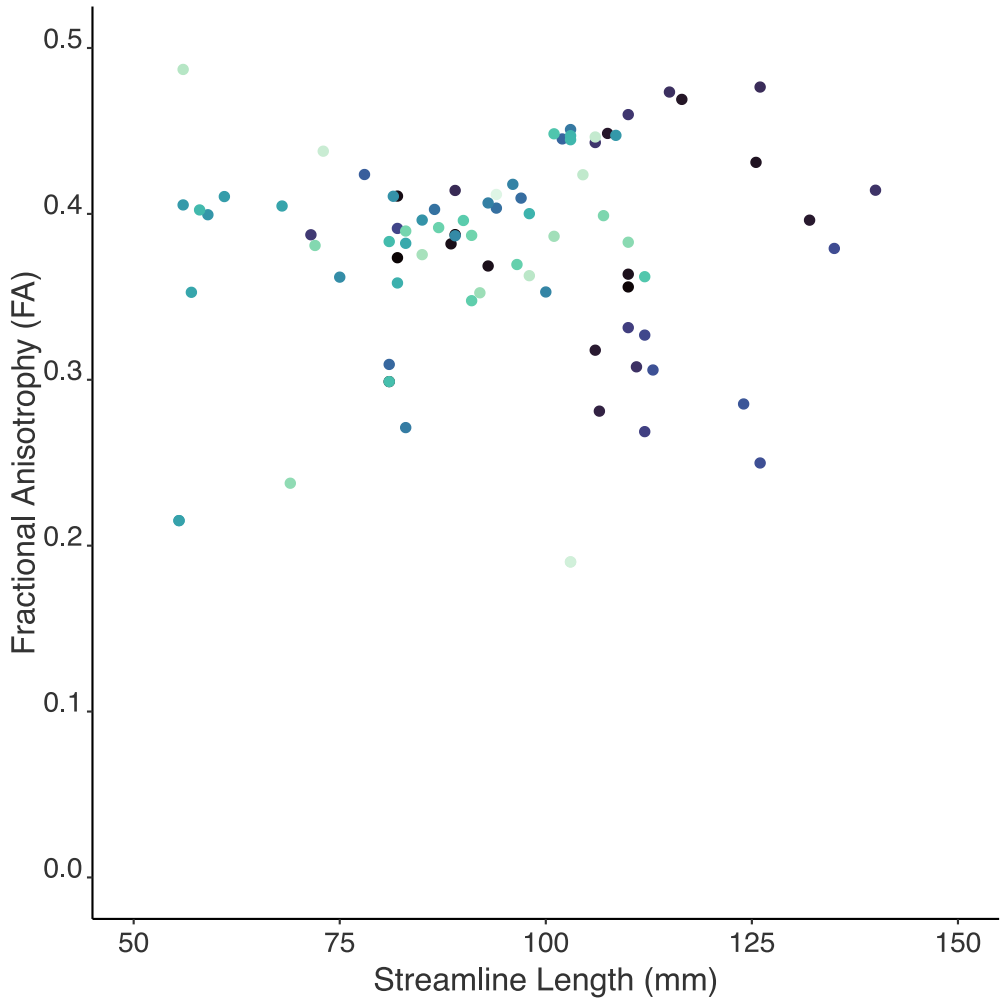

Figure 4

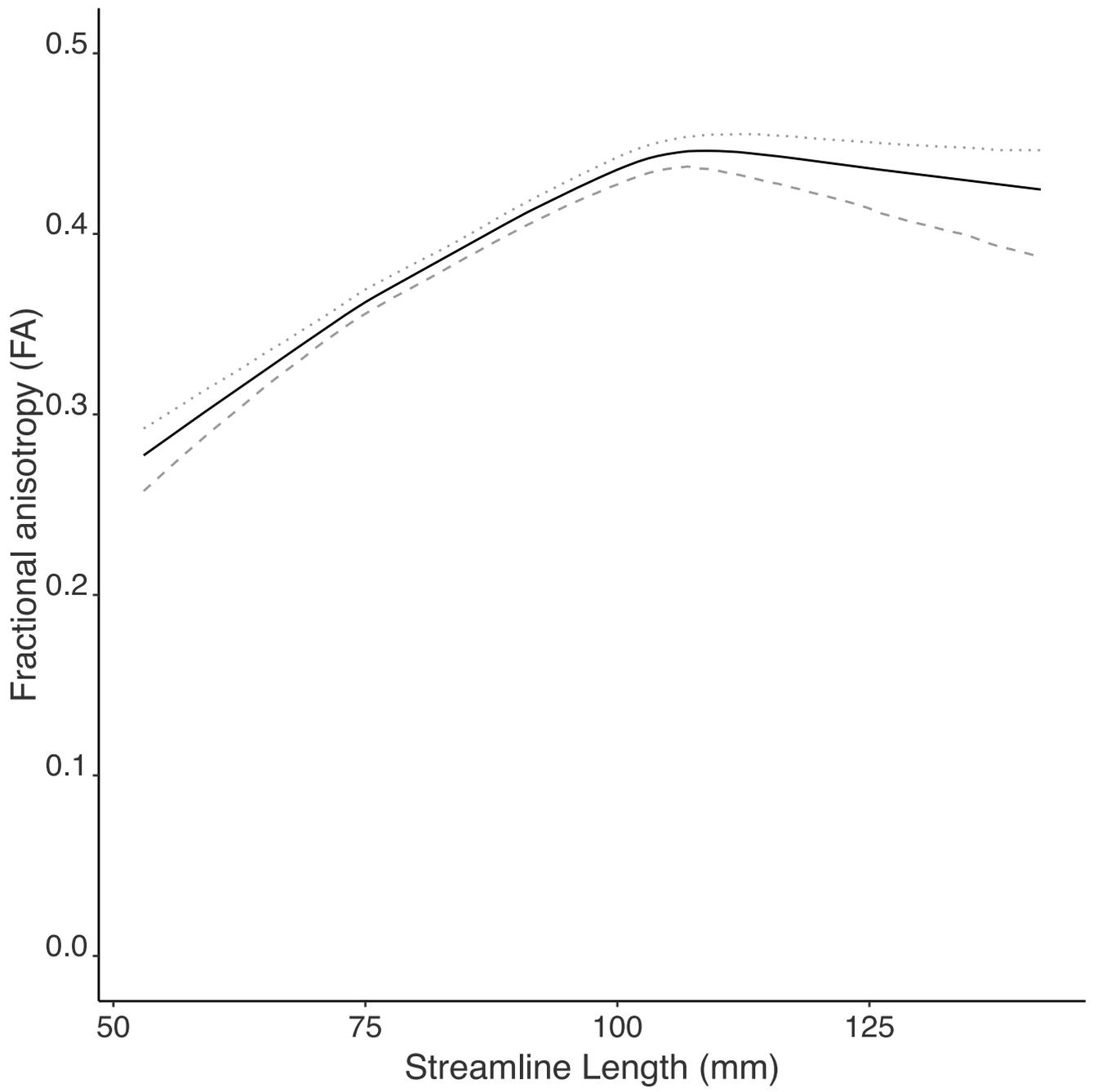

Figure 5

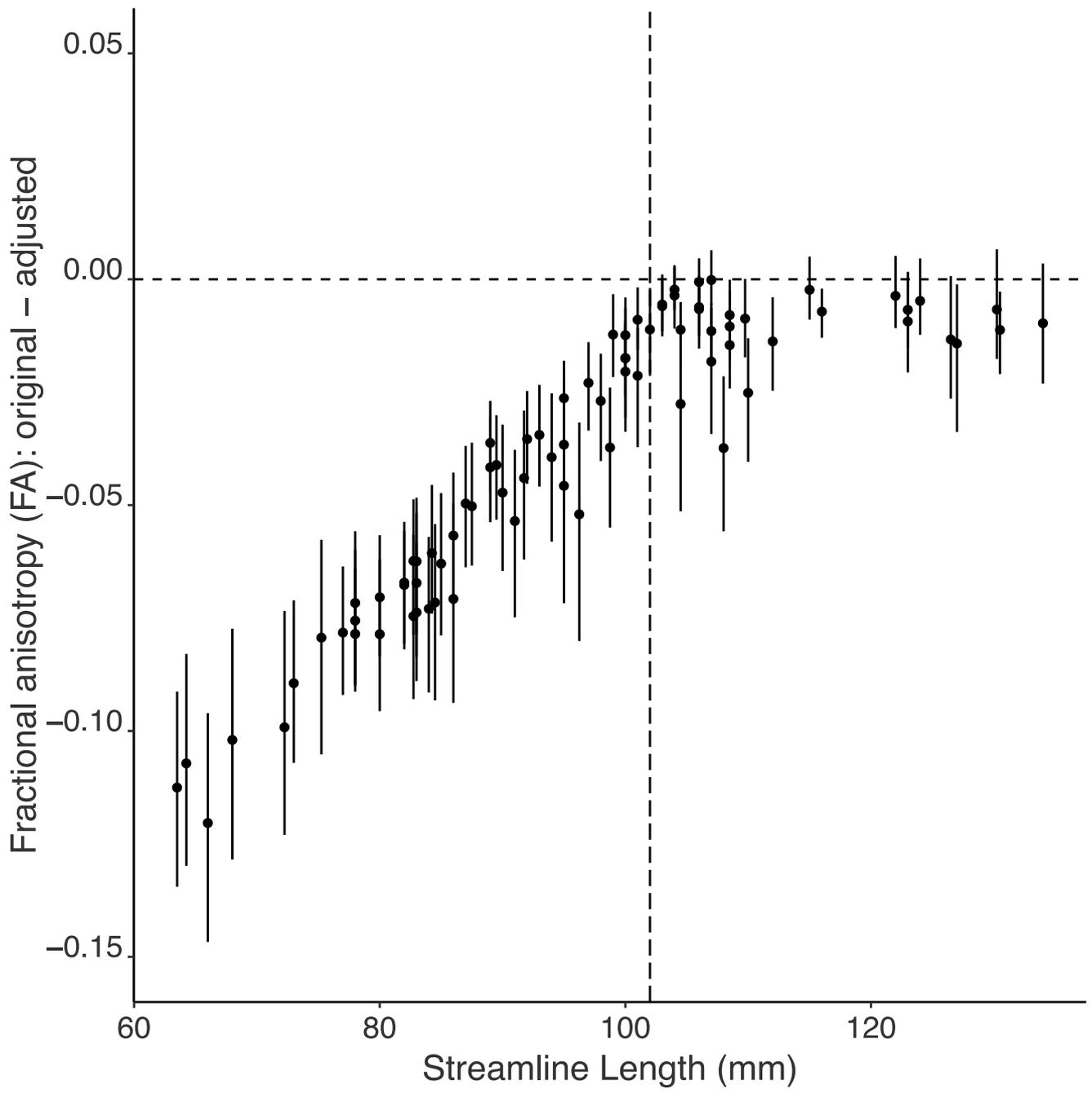

# Figure 6

A

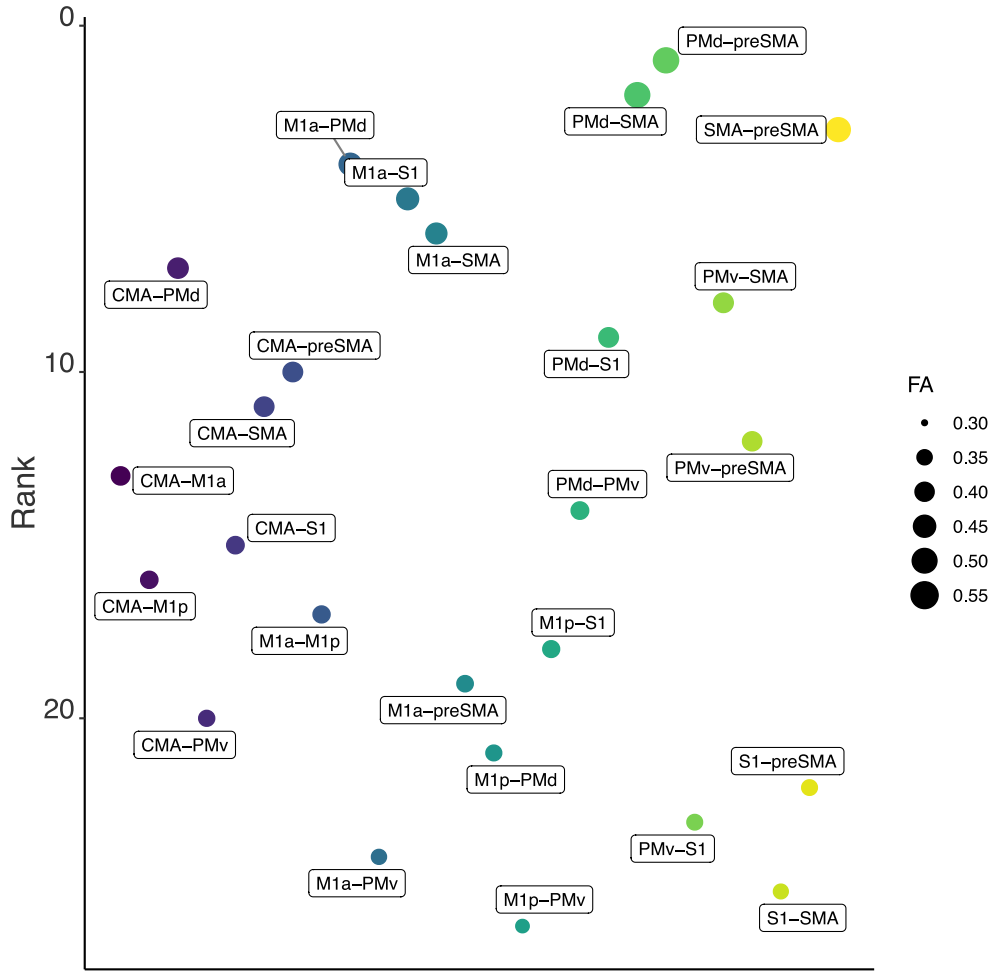

B

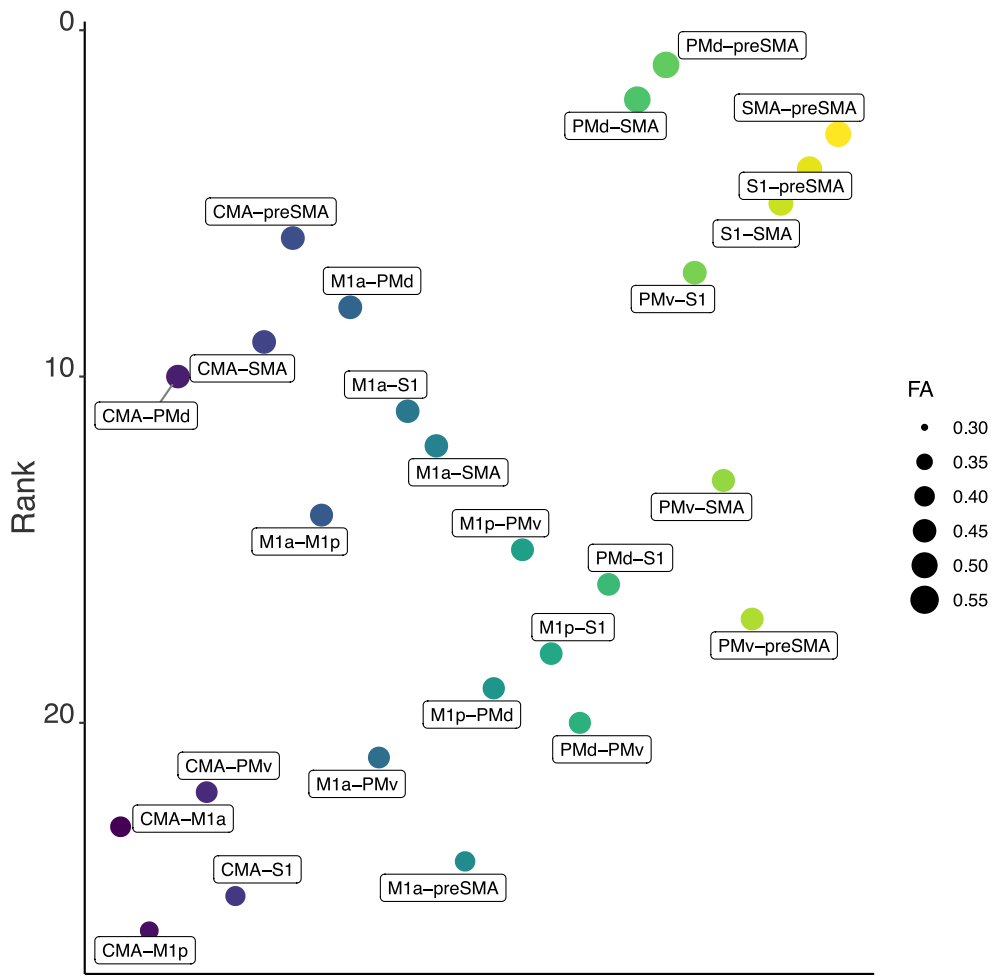

Figure 7

A

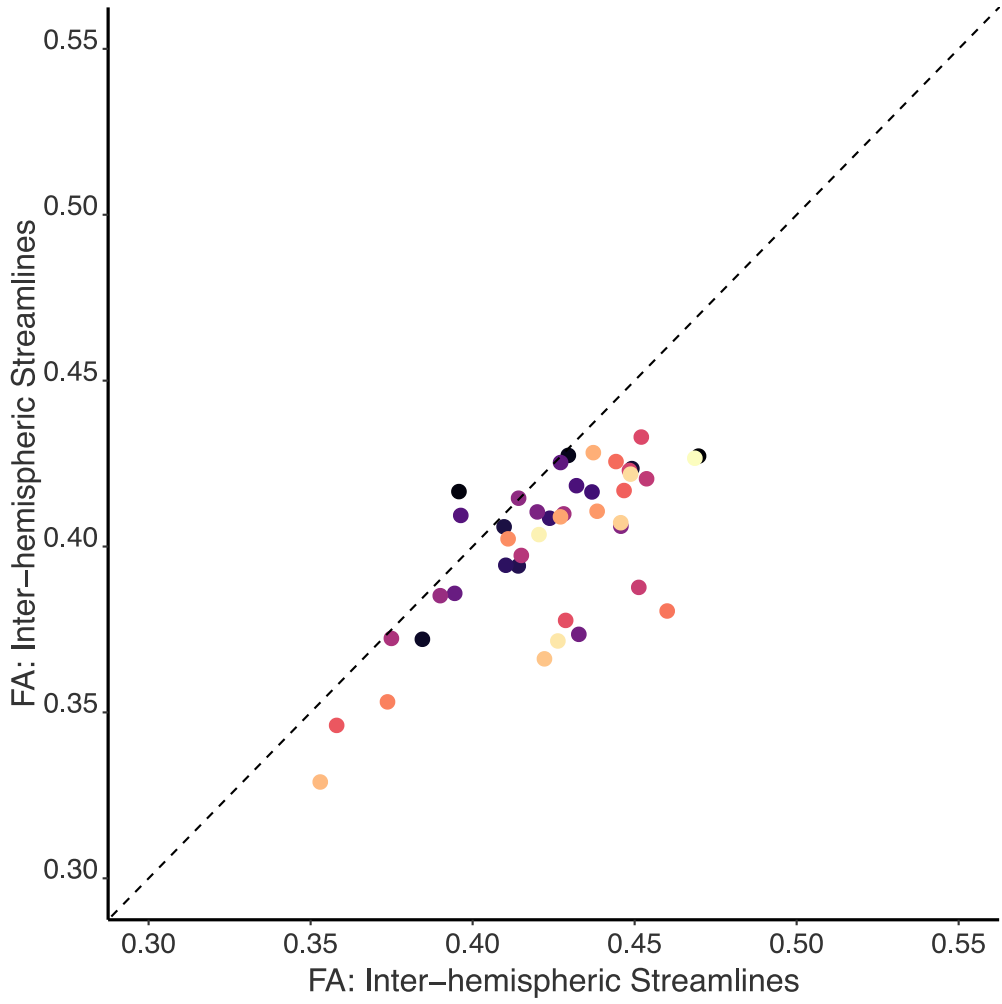

B

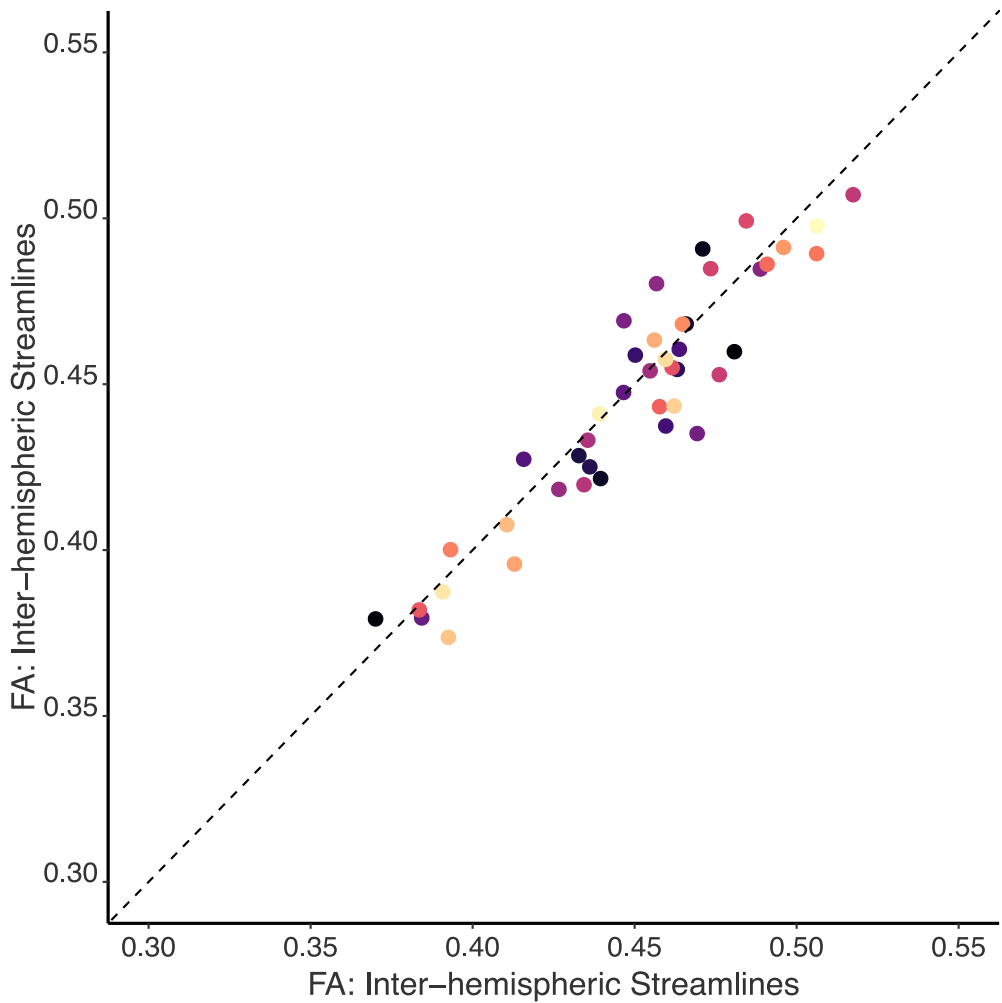

# Figure 8

A

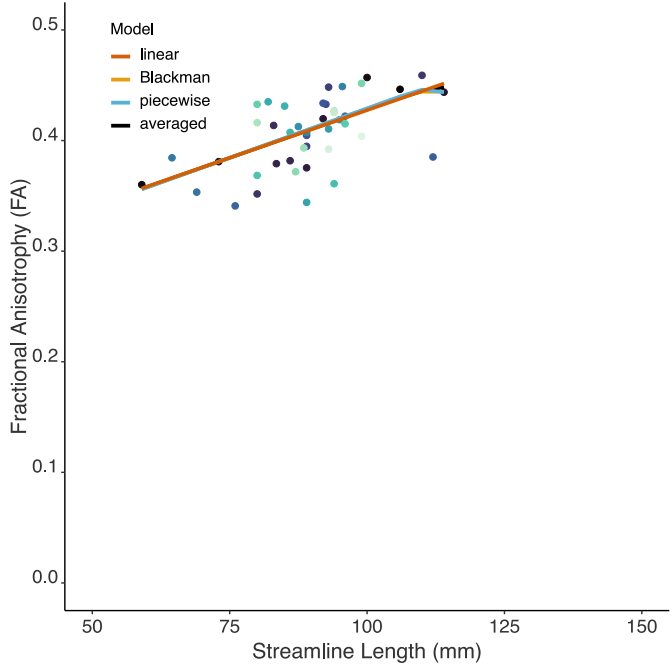

B

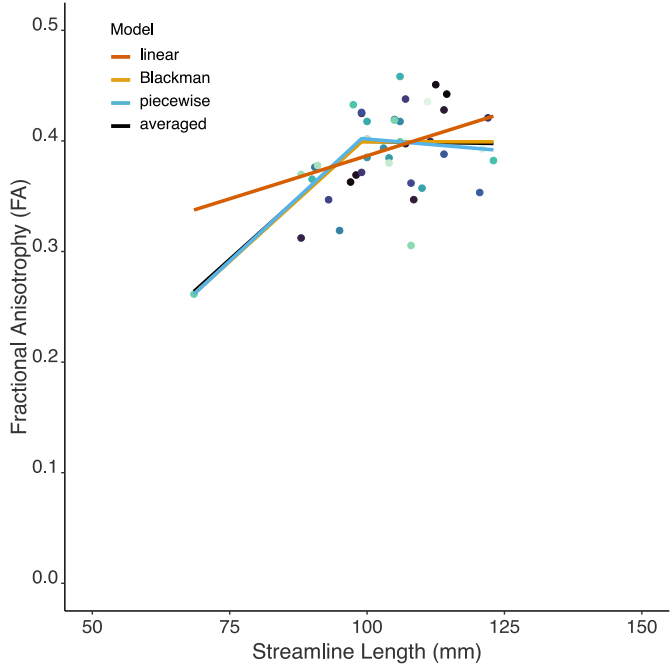

C

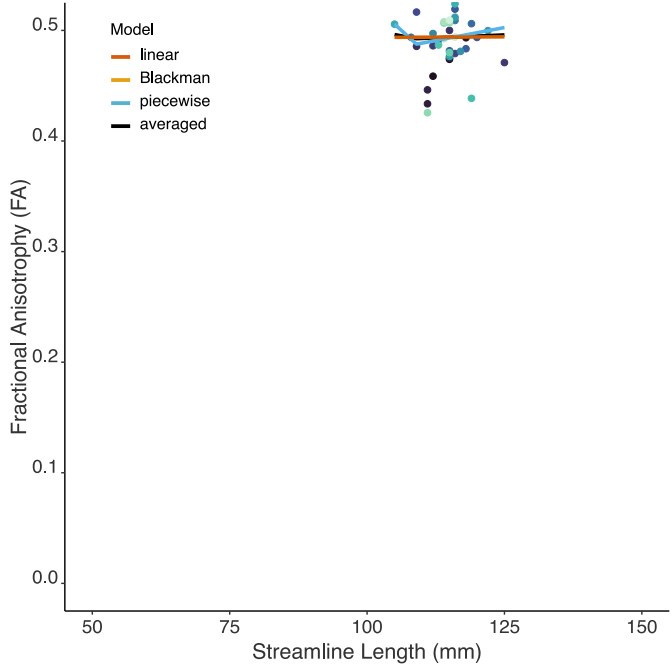

**Supplementary material**

Data Set 1

Further analyses using the methods detailed in the paper, were applied to the data described in Ruddy et al. (2017). These demonstrated that estimates of tissue microstructure other than fractional anisotropy (FA) are also dependent on streamline length. As an initial step, the Kendall rank correlation coefficient was used to characterise the ordinal association of streamline length and, respectively, apparent fibre density (AFD), mean diffusivity (MD) and radial diffusivity (RD). This was done in two ways. In the first, the coefficient was calculated separately for each of the 43 participants. The mean value of tau – the Kendall correlation coefficient, was then obtained (with corresponding confidence intervals) across participants (Table S1). Within individual brains, tracts with longer streamlines were characterised by larger AFD and MD values, and by smaller RD values.

**Table S1**

|  | lower c.i. | mean | upper c.i. |
|---|---|---|---|
| AFD vs. streamline length | 0.33 | 0.38 | 0.42 |
| MD vs. streamline length | 0.05 | 0.11 | 0.16 |
| RD vs. streamline length | -0.30 | -0.26 | -0.22 |

Kendall correlation coefficients (tau) which characterise the magnitude of the association between each estimate of tissue microstructure and streamline length.

In the second method of analysis, the coefficient was calculated separately for each of the tracts (e.g., left M1a to right M1a), using the sample of 43 participants. The mean value of tau across all tracts obtained (with corresponding confidence intervals) was then derived (Table S2). It is apparent that when any given tract is considered across different brains, individuals with longer streamlines tend to exhibit larger AFD and MD values, and smaller RD values.

**Table S2**

|                            | lower c.i. | mean  | upper c.i. |
|----------------------------|------------|-------|------------|
| AFD vs. streamline length  | 0.25       | 0.29  | 0.32       |
| MD vs. streamline length   | 0.02       | 0.05  | 0.08       |
| RD vs. streamline length   | -0.22      | -0.18 | -0.14      |

Kendall correlation coefficients (tau) which characterise the magnitude of the association between each estimate of tissue microstructure and streamline length.

Further analyses were then undertaken to determine for each estimate of tissue microstructure, the nature of the relationship with streamline length. Model averaged estimates (with confidence intervals) derived using linear, Blackman and piecewise linear models were obtained in the manner described in the paper.

Apparent fibre density (AFD)

In 26 of the 43 participants, the fit to the piecewise linear function was superior to that achieved using a Blackman (n = 15) or linear model (n = 2). The model averaged estimates (with 95% confidence intervals (c.i.)) are given in Table S3.

**Table S3**

|  | lower c.i. | estimate | upper c.i. |
|---|---|---|---|
| breakpoint (mm) | 99.4 | 103.2 | 107.2 |
| AFD at breakpoint | 0.76 | 0.79 | 0.82 |
| slope initial segment | 0.0075 | 0.0084 | 0.0093 |
| slope second segment | -0.0049 | -0.0026 | -0.0012 |
| tau pre-breakpoint | 0.48 | 0.52 | 0.55 |
| tau post-breakpoint | -0.16 | -0.09 | -0.03 |

A summary of the values obtained through the application of a model averaging approach to the parameter estimates generated by the three candidate models (linear, Blackman, and piecewise-linear). The lower and upper 95% confidence intervals were generated from 1000 bootstrapped samples (drawn from the cohort of forty-three participants). The slope values represent the change in AFD with respect to each 1 mm increase in streamline length, for i) the initial segment, ii) the second segment. The tau values are the Kendall correlation coefficients for these segments. These characterise the magnitude of the association between AFD and streamline length.

The mean magnitude of the Kendall correlation between AFD values adjusted for the influence of streamline length (i.e., through the application of a model averaging approach) and streamline length, when calculated across participants, was -0.021 (95% c.i. -0.038 – -0.006). Calculated across tracts, the correlation between the (streamline length) adjusted AFD and streamline length did not differ reliably from zero (mean = -0.023, 95% c.i. -0.067 – 0.020).

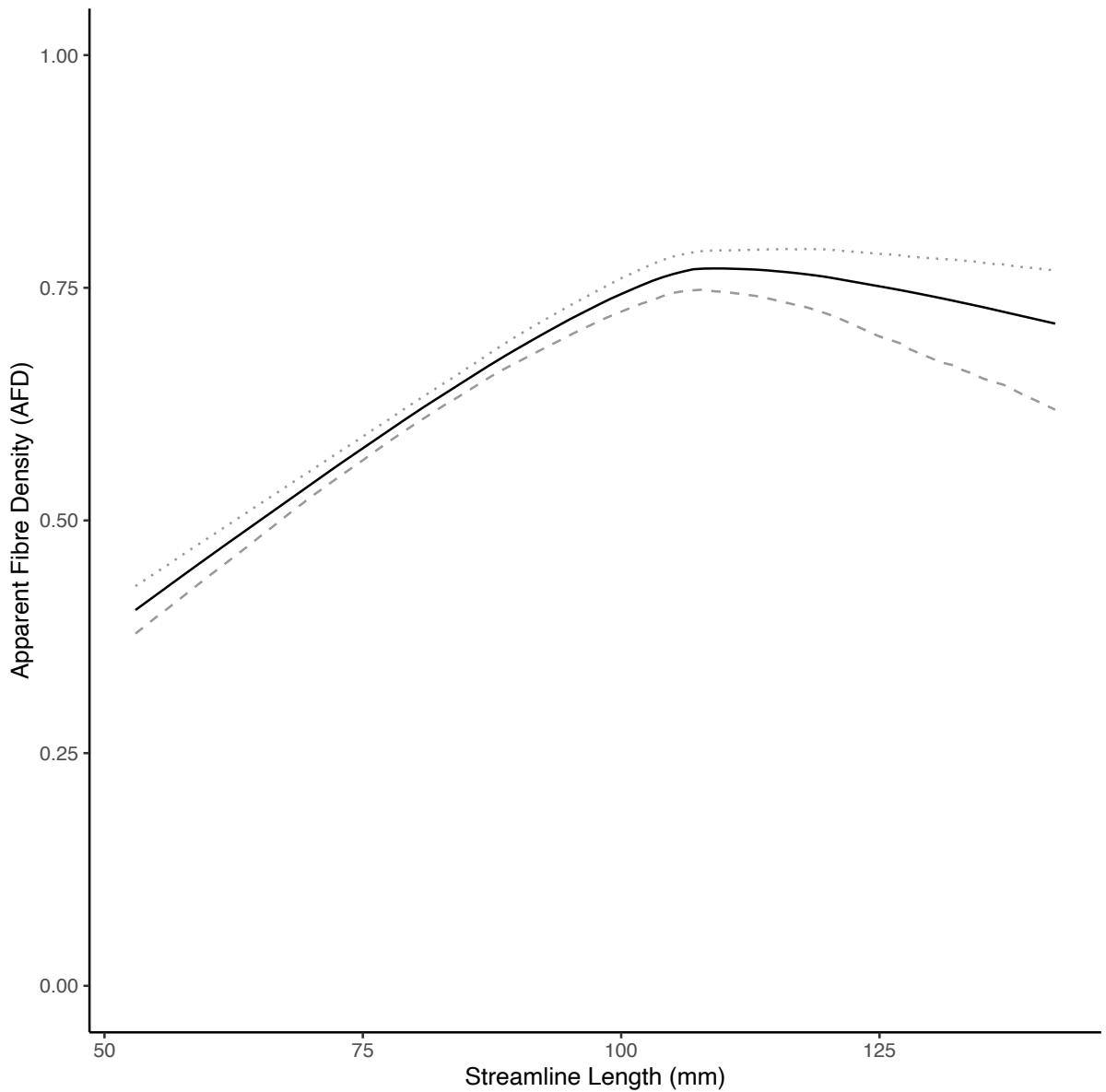

Figure S1. A summary representation of the results obtained through the application of a model averaging approach to the predicted AFD values generated by the three candidate models (linear, Blackman, and piecewise-linear), for the 43 participants included in Ruddy et al. (2017). For each participant, the models were evaluated, and the predictions weighted and averaged, at a range of nominal streamline lengths (at 1mm intervals). This range spanned the median minimum streamline length and the median maximum streamline length observed across the 43 participants. The solid line corresponds to the means of the weighted, averaged, predicted values derived from 1000 bootstrapped samples. The dashed line was generated using the lower 95% confidence interval of the bootstrapped samples. The dotted line was generated using the upper 95% confidence interval. It is apparent that, for streamlines shorter than approximately 100 mm, there is a clear dependence of AFD on streamline length.

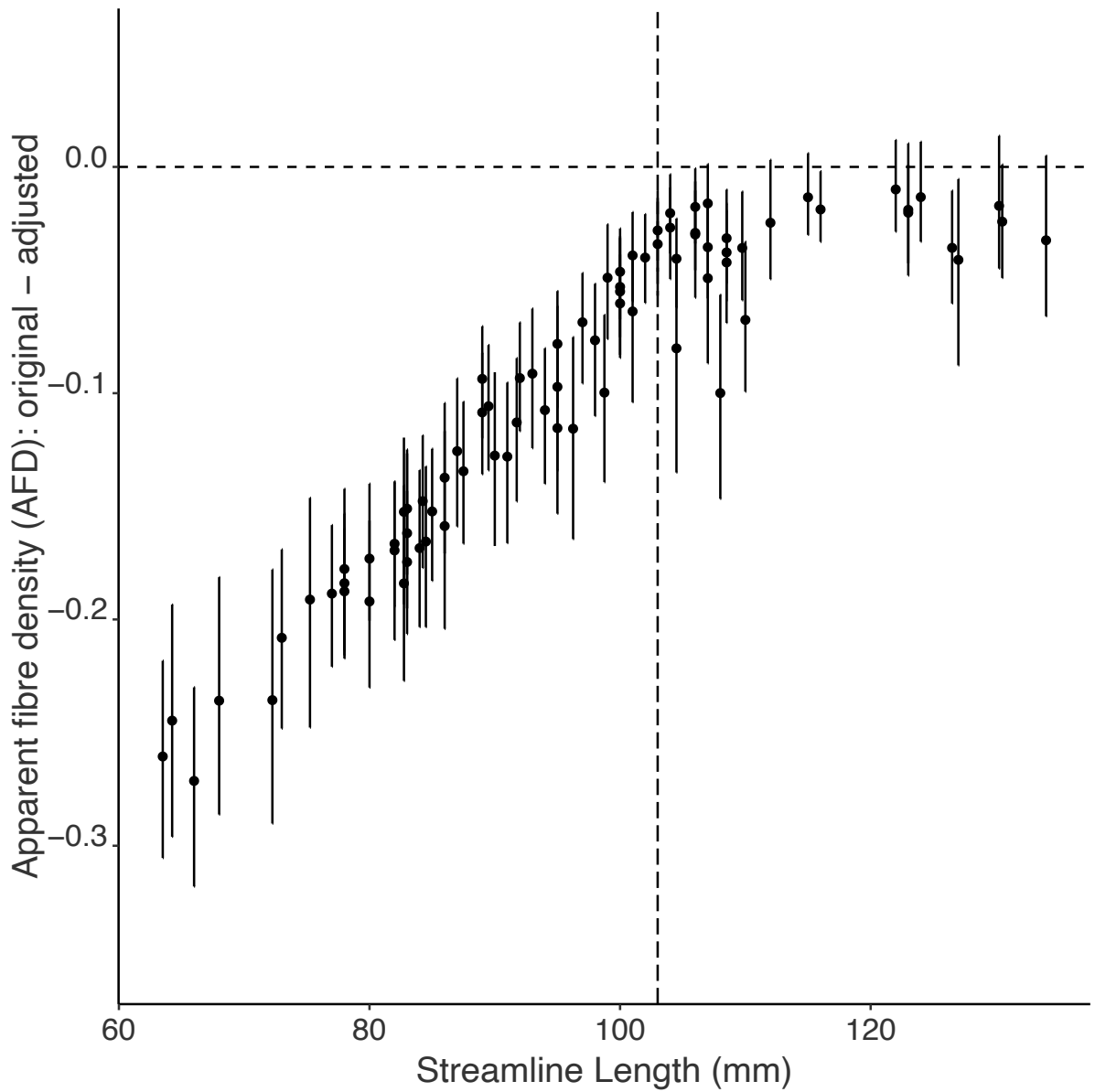

Figure S2. Separately for each of the 43 participants included in Ruddy et al. (2017), and for each tract, the difference between the original AFD value, and the AFD value adjusted for the influence of streamline length (through the application of a model averaging approach) was calculated. The filled symbols correspond to the means of these difference values, when derived from 1000 bootstrapped samples drawn from the set of 43 participants (i.e., calculated separately for each tract). For a tract to be included, it was necessary that at least half of the participants must have contributed data (i.e., streamlines were resolved). The error bars correspond to the 95% confidence intervals of the bootstrapped samples. The tracts are plotted in order of mean streamline length (calculated across participants).

Mean diffusivity (MD)

In 28 of the 43 participants, the fit to the piecewise linear function was superior to that achieved using a Blackman (n = 3) or linear model (n = 12). The model averaged estimates (with 95% confidence intervals (c.i.)) are given in Table S4.

**Table S4**

|  | lower c.i. | estimate | upper c.i. |
|---|---|---|---|
| breakpoint (mm) | 88.1 | 105.4 | 123.6 |
| MD at breakpoint | 6.90E-04 | 6.96E-04 | 7.00E-04 |
| slope initial segment | -3.97E-06 | -1.67E-06 | -2.76E-07 |
| slope second segment | 2.08E-07 | 4.32E-07 | 7.77E-07 |
| tau pre-breakpoint | -0.01 | 0.07 | 0.14 |
| tau post-breakpoint | 0.05 | 0.12 | 0.18 |

A summary of the values obtained through the application of a model averaging approach to the parameter estimates generated by the three candidate models (linear, Blackman, and piecewise-linear). The lower and upper 95% confidence intervals were generated from 1000 bootstrapped samples (drawn from the cohort of forty-three participants). The slope values represent the change in MD with respect to each 1 mm increase in streamline length, for i) the initial segment, ii) the second segment. The tau values are the Kendall correlation coefficients for these segments. These characterise the magnitude of the association between MD and streamline length.

When calculated across participants, the mean magnitude of the Kendall correlation between the MD values adjusted for the influence of streamline length (i.e., through the application of a model averaging approach) and streamline length did not differ reliably from zero (mean = 0.002, 95% c.i.  -0.015 – 0.017). The mean magnitude of the Kendall correlation between the (streamline length) adjusted MD values and streamline length, when calculated across tracts, was 0.045 (95% c.i. 0.018 – 0.068).

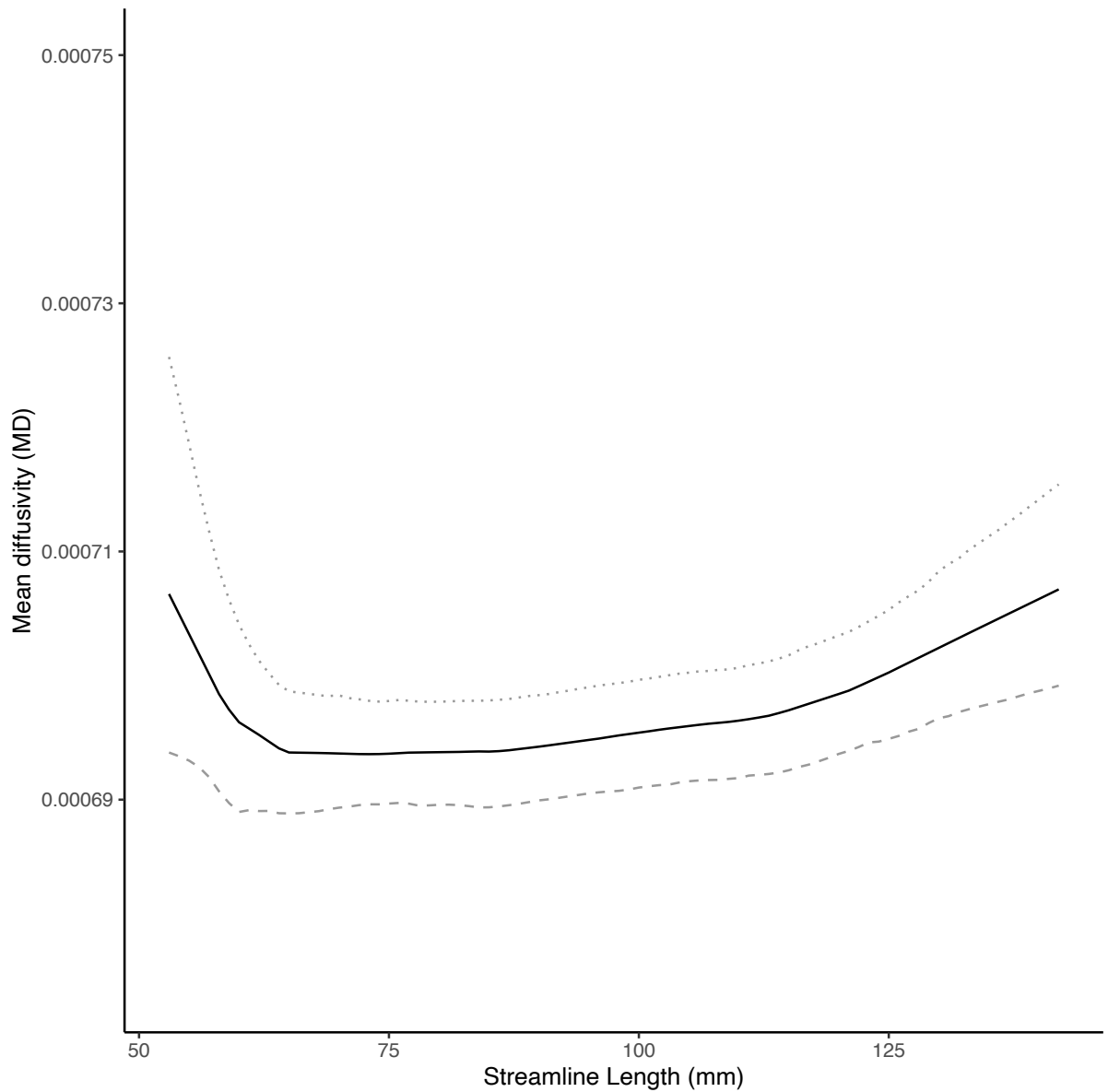

Figure S3. A summary representation of the results obtained through the application of a model averaging approach to the predicted MD values generated by the three candidate models (linear, Blackman, and piecewise-linear), for the 43 participants included in Ruddy et al. (2017). For each participant, the models were evaluated, and the predictions weighted and averaged, at a range of nominal streamline lengths (at 1mm intervals). This range spanned the median minimum streamline length and the median maximum streamline length observed across the 43 participants. The solid line corresponds to the means of the weighted, averaged, predicted values derived from 1000 bootstrapped samples. The dashed line was generated using the lower 95% confidence interval of the bootstrapped samples. The dotted line was generated using the upper 95% confidence interval.

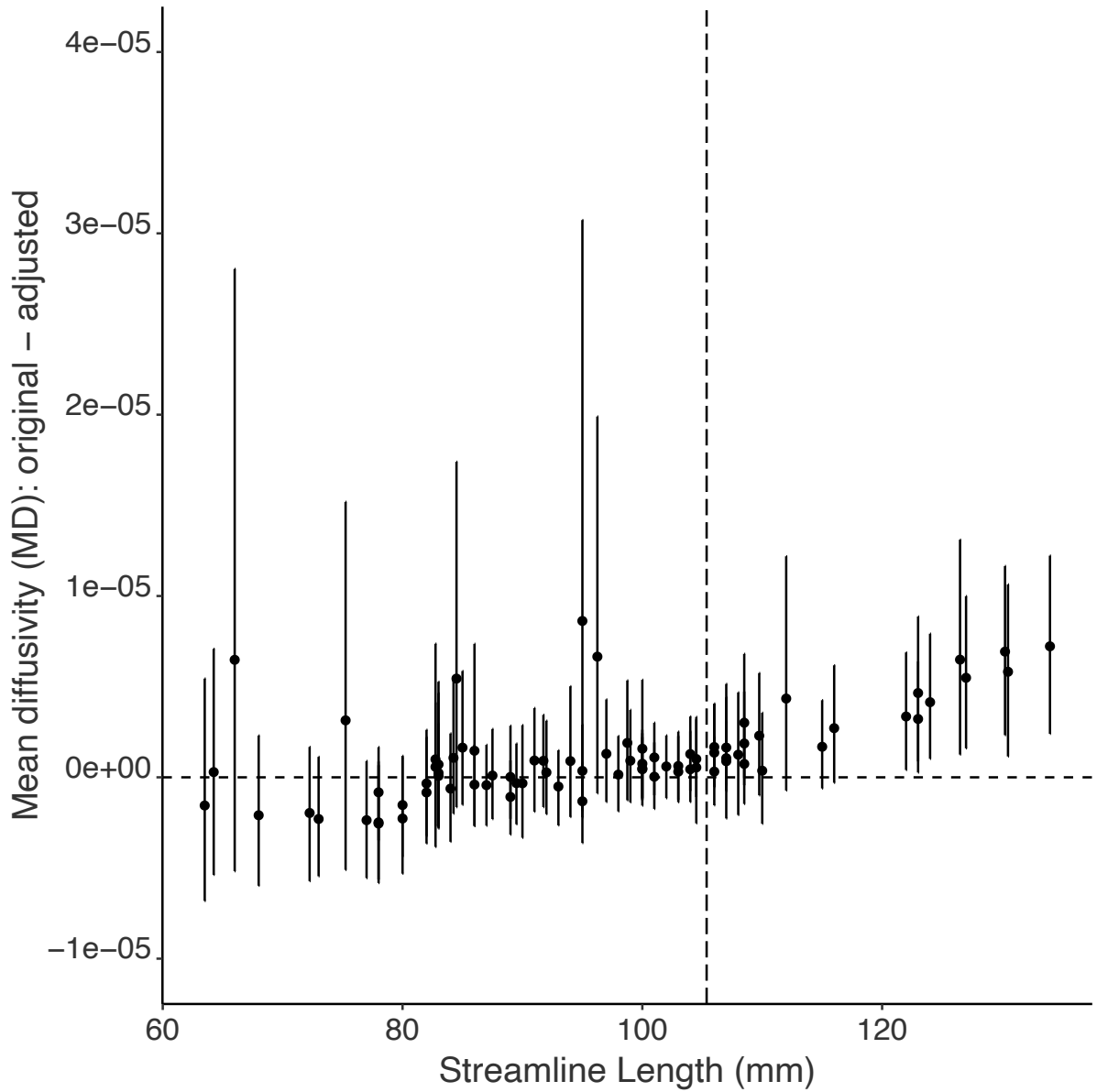

Figure S4. Separately for each of the 43 participants included in Ruddy et al. (2017), and for each tract, the difference between the original MD value, and the MD value adjusted for the influence of streamline length (through the application of a model averaging approach) was calculated. The filled symbols correspond to the means of these difference values, when derived from 1000 bootstrapped samples drawn from the set of 43 participants (i.e., calculated separately for each tract). For a tract to be included, it was necessary that at least half of the participants must have contributed data (i.e., streamlines were resolved). The error bars correspond to the 95% confidence intervals of the bootstrapped samples. The tracts are plotted in order of mean streamline length (calculated across participants).

Radial diffusivity (RD)

In 29 of the 43 participants, the fit to the piecewise linear function was superior to that achieved using a Blackman (n = 8) or linear model (n = 6). The model averaged estimates (with 95% confidence intervals (c.i.)) are given in Table S5.

**Table S5**

|  | lower c.i. | estimate | upper c.i. |
|---|---|---|---|
| breakpoint (mm) | 99.2 | 103.6 | 107.4 |
| RD at breakpoint | 5.01E-04 | 5.08E-04 | 5.16E-04 |
| slope initial segment | -4.38E-06 | -2.68E-06 | -1.49E-06 |
| slope second segment | 4.83E-07 | 9.41E-07 | 1.49E-06 |
| tau pre-breakpoint | -0.46 | -0.42 | -0.36 |
| tau post-breakpoint | 0.05 | 0.15 | 0.22 |

A summary of the values obtained through the application of a model averaging approach to the parameter estimates generated by the three candidate models (linear, Blackman, and piecewise-linear). The lower and upper 95% confidence intervals were generated from 1000 bootstrapped samples (drawn from the cohort of forty-three participants). The slope values represent the change in RD with respect to each 1 mm increase in streamline length, for i) the initial segment, ii) the second segment. The tau values are the Kendall correlation coefficients for these segments. These characterise the magnitude of the association between RD and streamline length.

The mean magnitude of the Kendall correlation between RD values adjusted for the influence of streamline length (i.e., through the application of a model averaging approach) and streamline length, did not differ reliably from zero (mean = -0.010, 95% c.i.  -0.029 – 0.009). Likewise, when calculated across tracts, the mean magnitude of the Kendall correlation between the (streamline length) adjusted RD values and streamline length did not differ reliably from zero (mean = 0.017, 95% c.i.  -0.020 – 0.056)

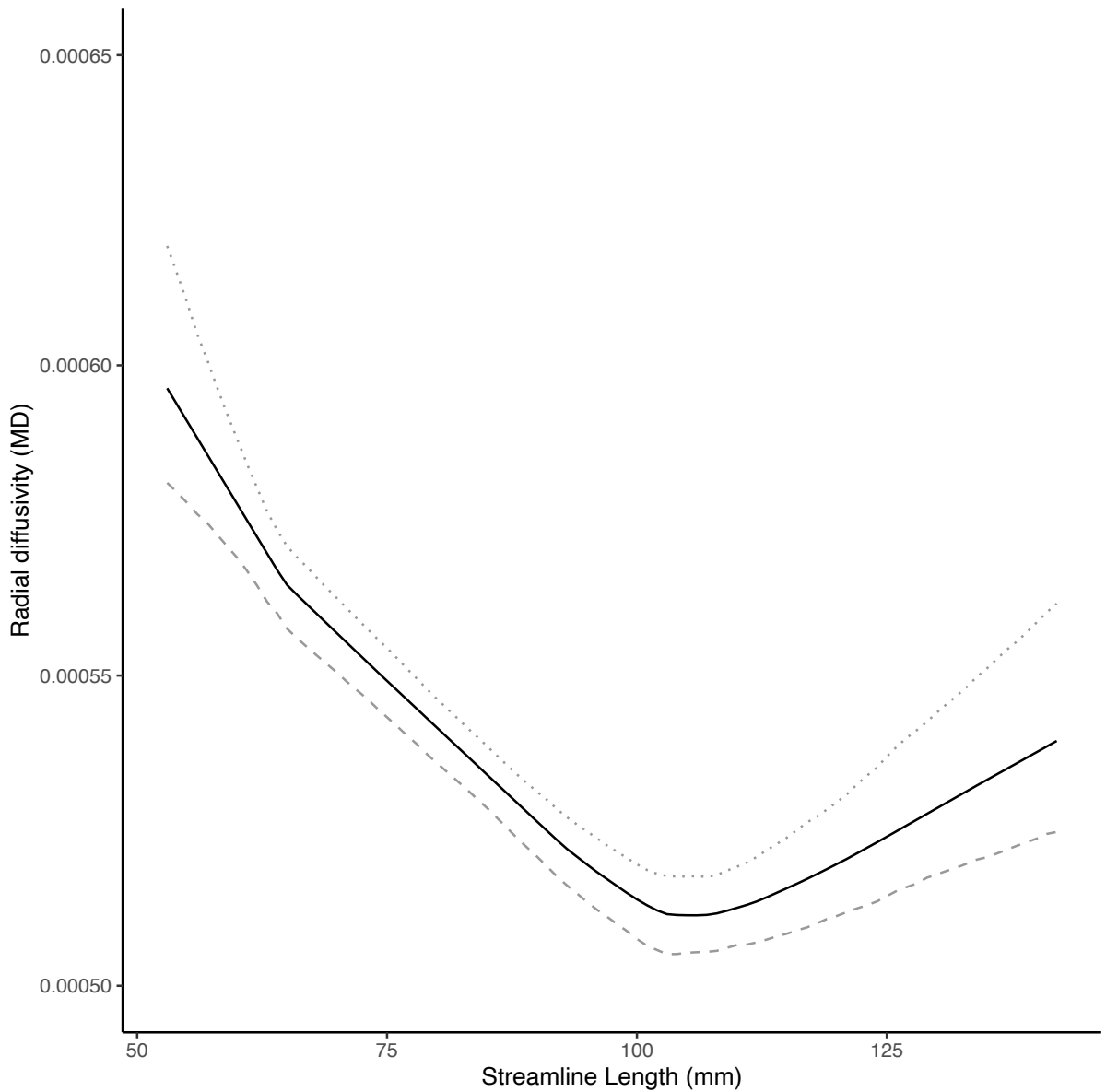

Figure S5. A summary representation of the results obtained through the application of a model averaging approach to the predicted RD values generated by the three candidate models (linear, Blackman, and piecewise-linear), for the 43 participants included in Ruddy et al. (2017). For each participant, the models were evaluated, and the predictions weighted and averaged, at a range of nominal streamline lengths (at 1mm intervals). This range spanned the median minimum streamline length and the median maximum streamline length observed across the 43 participants. The solid line corresponds to the means of the weighted, averaged, predicted values derived from 1000 bootstrapped samples. The dashed line was generated using the lower 95% confidence interval of the bootstrapped samples. The dotted line was generated using the upper 95% confidence interval.

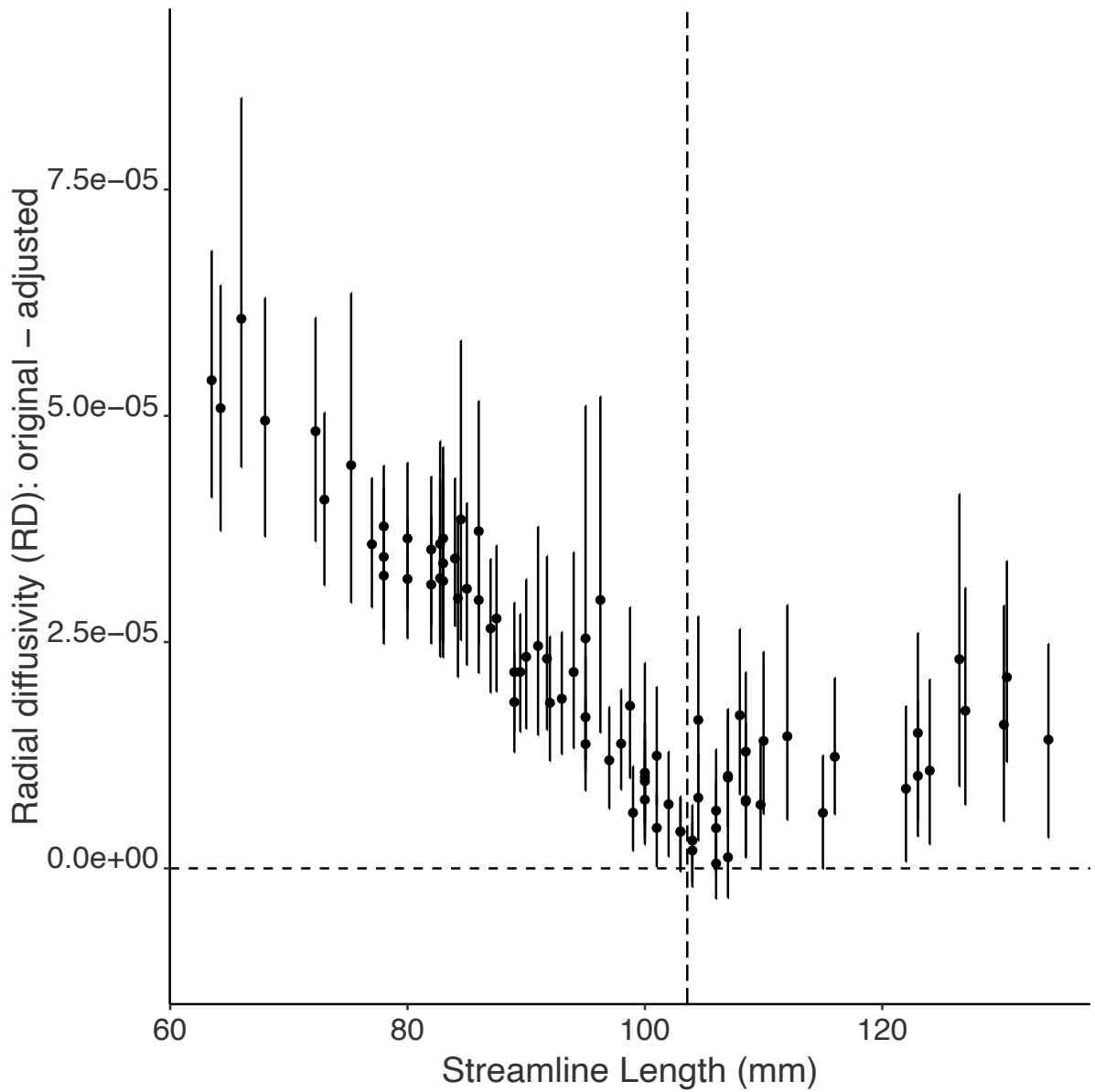

Figure S6. Separately for each of the 43 participants included in Ruddy et al. (2017), and for each tract, the difference between the original RD value, and the RD value adjusted for the influence of streamline length (through the application of a model averaging approach) was calculated. The filled symbols correspond to the means of these difference values, when derived from 1000 bootstrapped samples drawn from the set of 43 participants (i.e., calculated separately for each tract). For a tract to be included, it was necessary that at least half of the participants must have contributed data (i.e., streamlines were resolved). The error bars correspond to the 95% confidence intervals of the bootstrapped samples. The tracts are plotted in order of mean streamline length (calculated across participants).

Data Set 2

To demonstrate that a dependence of estimates of tissue microstructure on streamline length is observed more generally, the analysis of a different data set is reported. The data were obtained in the context of a study described in Hänggi et al. (2017) and Saettaet al. (2022). The participants were sixteen neurologically healthy males (aged $44.18 \pm 9.51$ SD). All gave informed consent to procedures that were in accordance with the Declaration of Helsinki. These had been approved by Ethics Committee of the University Hospital of Zurich and by the Ethical Committee Milano Area C. Eight participants were enrolled in Zurich, and eight in Milan.

At the University Hospital of Zurich, diffusion weighted imaging (DWI) was undertaken using a Philips Achieva 3.0 Tesla whole-body scanner (Philips Medical Systems, Best, The Netherlands) was used in conjunction with an eight-element sensitivity encoding (SENSE) head coil array. The DWI sequence was characterised by a repetition time of $13.010$ s, echo time = $55$ ms, b value = $1000$ s/mm$^2$, flip angle = $90°$, spatial resolution = $2 \times 2 \times 2$ mm$^3$ (matrix $112 \times 112$ pixels, number of slices = $75$ (transverse)), field of view = $224 \times 224$ mm$^2$, SENSE factor, $2.1$. Diffusion was registered in $32$ noncollinear directions. A non-diffusion weighted reference volume (b = $0$ s/ mm$^2$) was also first obtained. At the Neuroradiology Department of the "ASST Grande Ospedale Metropolitano Niguarda" of Milan, the data were acquired using a General Electric (GE) 1.5 Tesla Signa HD-XT scanner. The DWI sequence was characterised by a b value of $700$ s/mm$^2$. Diffusion was registered in $35$ noncollinear directions, with a spatial resolution of $1.094 \times 1.094 \times 4$ mm$^3$ . A preceding non-diffusion weighted reference volume (b = $0$ s/ mm$^2$) was obtained.

ExploreDTI (Leemans et al. 2009) was used for data processing. Images were corrected for head movement and eddy currents using the procedure described in Leemans and Jones (2009). Diffusion tensor estimation was performed using the iteratively reweighted linear least squares approach (Veraart et al. 2013). Whole-brain deterministic tractography was undertaken with a seed point resolution of 2 x 2 x 2 mm and a Fractional Anisotropy (FA) threshold of 0.2.

Reconstructed fibre trajectories were generated for all pairwise combinations of brain regions of interest (RoI) defined within the automated anatomical labelling (AAL) atlas (Rolls et al., 2015). This encompasses 90 RoIs for each hemisphere, including the whole cerebrum and subcortical structures, but excluding the cerebellum. In total, 4005 unique tracts (within and between hemispheres) are thus defined. The median number of tracts for which streamlines were present (i.e., across the sixteen participants) was 1342.2 (95% c.i. 1219.9 – 1461.2). For every such instance, estimates of fractional anisotropy (FA) and streamline length were derived. The relationship between FA and streamline length was analysed separately for each participant in the manner described in the main text.

In 11 of the 16 participants, the fit to the piecewise linear function was superior to that achieved using a Blackman model (n = 5). In no instance was the best fit achieved using a linear model.  A summary of the fitted parameters obtained for the cohort of sixteen participants is provided in Table S6. The mean range of streamline lengths included in the analyses (i.e., across the sixteen participants) was 300.1 mm (95% c.i. 258.9 – 345.7 mm). A summary representation of the results obtained through the application of a model averaging approach to the to the predicted FA values generated by the three candidate models (linear, Blackman, and piecewise-linear) is shown in Figure S7.

**Table S6**

|  | lower c.i. | estimate | upper c.i. |
|---|---|---|---|
| breakpoint (mm) | 111.4 | 123.5 | 144.6 |
| FA at breakpoint | 0.46 | 0.47 | 0.48 |
| slope initial segment | 0.0013 | 0.0015 | 0.0017 |
| slope second segment | -1.87E-04 | -1.98E-05 | 1.22E-04 |
| tau pre-breakpoint | 0.35 | 0.37 | 0.40 |
| tau post-breakpoint | 0.03 | 0.12 | 0.15 |

A summary of the values obtained through the application of a model averaging approach to the parameter estimates generated by the three candidate models (linear, Blackman, and piecewise-linear). The lower and upper 95% confidence intervals were generated from 1000 bootstrapped samples (drawn from the cohort of sixteen participants). The slope values represent the change in FA with respect to each 1 mm increase in streamline length, for i) the initial segment, ii) the second segment. The tau values are the Kendall correlation coefficients for these segments. These characterise the magnitude of the association between FA and streamline length.

**Figure S7**

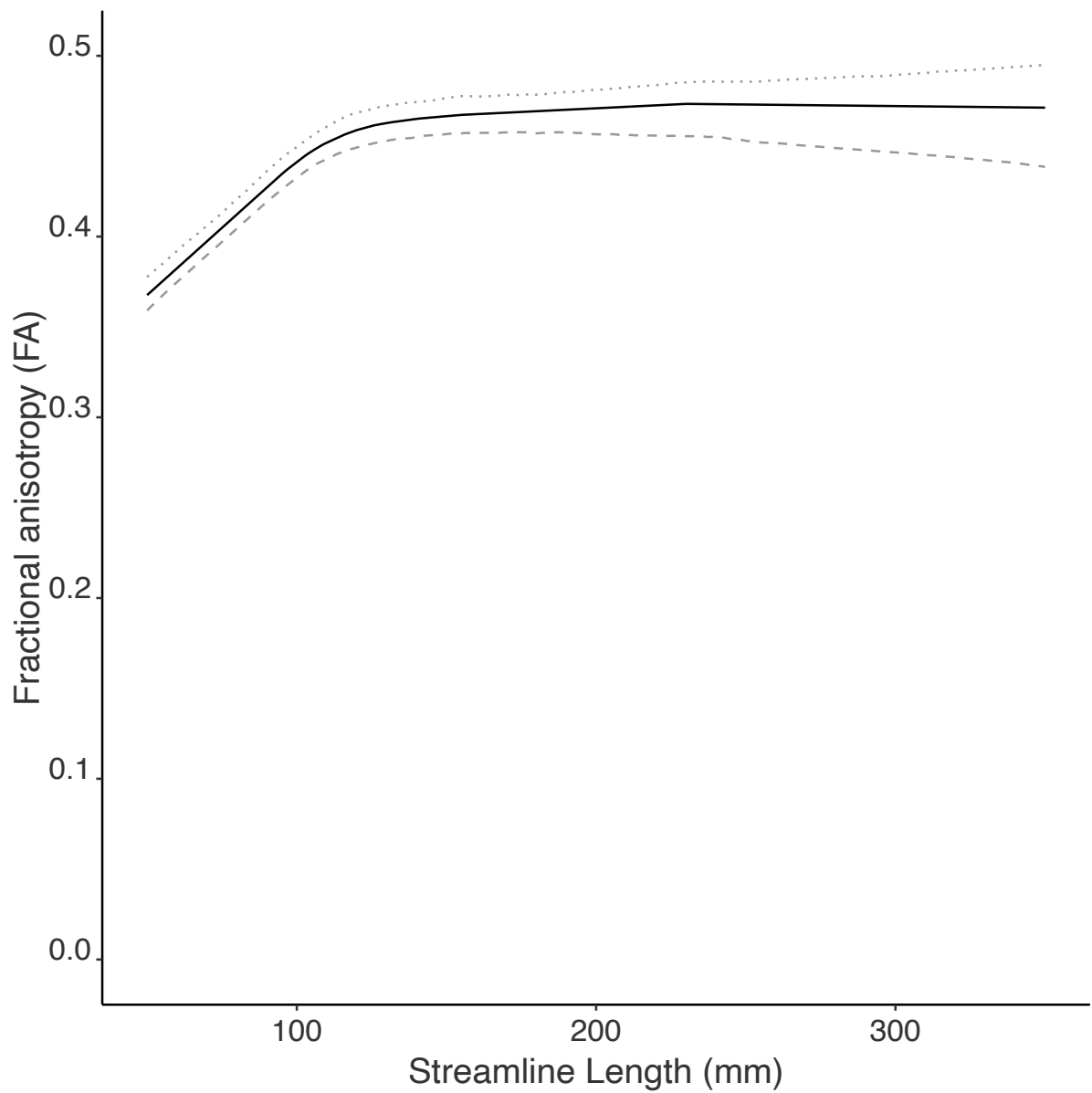

A summary representation of the results obtained through the application of a model averaging approach to the predicted FA values generated by the three candidate models (linear, Blackman, and piecewise-linear) for the 16 participants. For each participant, the models were evaluated, and the predictions weighted and averaged, at a range of nominal streamline lengths (at 1mm intervals). This range spanned the mean minimum streamline length (50 mm) and the mean maximum streamline length (346 mm) observed across the 16 participants. The solid line corresponds to the means of the weighted, averaged, predicted values derived from 1000 bootstrapped samples. The dashed line was generated using the lower 95% confidence interval of the bootstrapped samples. The dotted line was generated using the upper 95% confidence interval.